\documentclass[journal,10pt]{IEEEtran}
\usepackage[cmex10]{amsmath}
\usepackage{graphicx}
\usepackage{algorithm}
\usepackage{algorithmic}
\usepackage{stfloats}
\usepackage{epstopdf}
\usepackage[numbers,sort&compress]{natbib}
\usepackage{multirow}
\usepackage{color}
\usepackage{balance}
\usepackage{amsmath, amssymb}
\usepackage{amsfonts}
\usepackage{caption}
\usepackage{subfigure}
\usepackage{hyperref}
\usepackage{comment}

\usepackage{bm}
\ifCLASSINFOpdf
\else
\fi
\begin{document}
\title{STAR-RIS-Enabled Full-Duplex Integrated Sensing and Communication System}
\author{Yu~Liu,~\IEEEmembership{Student Member,~IEEE,}~Gaojie~Chen,~\IEEEmembership{Senior Member,~IEEE,}
Yun~Wen,~\IEEEmembership{Student Member,~IEEE,}\\
Qu~Luo,~\IEEEmembership{Member,~IEEE,}
Chiya~Zhang,~\IEEEmembership{Member,~IEEE,}
and Dusit~Niyato,~\IEEEmembership{Fellow,~IEEE}

\thanks{\noindent Yu Liu, Gaojie Chen, Yun Wen, and Qu Luo are with the Institute for Communication Systems (ICS), 5GIC \& 6GIC, University of Surrey, Guildford, Surrey GU2 7XH, U.K. (e-mail: \{y.u.liu, gaojie.chen, yun.wen, q.u.luo\}@surrey.ac.uk. (\textit{Corresponding author: Gaojie Chen.})

Chiya Zhang is with the School of Electronic and Information Engineering,
Harbin Institute of Technology, Shenzhen 518055, China, also with the Peng
Cheng Laboratory (PCL), Shenzhen 5180523, China, and also with the National
Mobile Communications Research Laboratory, Southeast University, Nanjing
211189, China (e-mail: zhangchiya@hit.edu.cn).

Dusit Niyato is with the School of Computer Science and Engineering, Nanyang
Technological University, Singapore (e-mail: dniyato@ntu.edu.sg).}

}

\maketitle

\begin{abstract} 
Traditional self-interference cancellation (SIC) methods are common in full-duplex (FD) integrated sensing and communication (ISAC) systems. However, exploring new SIC schemes is important due to the limitations of traditional approaches. 
With the challenging limitations of traditional SIC approaches, this paper proposes a novel simultaneous transmitting and reflecting reconfigurable intelligent surface (STAR-RIS)-enabled FD ISAC system, where STAR-RIS enhances simultaneous communication and target sensing and reduces self-interference (SI) to a level comparable to traditional SIC approaches. The optimization of maximizing the sensing signal-to-interference-plus-noise ratio (SINR) and the communication sum rate, both crucial for improving sensing accuracy and overall communication performance, presents significant challenges due to the non-convex nature of these problems.
Therefore, we develop alternating optimization algorithms to iteratively tackle these problems. Specifically, we devise the semi-definite relaxation (SDR)-based algorithm for transmit beamformer design. For the reflecting and refracting coefficients design, we adopt the successive convex approximation (SCA) method and implement the SDR-based algorithm to tackle the quartic and quadratic constraints. 
Simulation results validate the effectiveness of the proposed
algorithms and show that the proposed deployment can achieve
better performance than that of the benchmark using the traditional SIC approach without STAR-RIS deployment.
\end{abstract}


\IEEEpeerreviewmaketitle

\section{Introduction}

The sixth generation (6G) wireless communication networks are anticipated to deliver high data rates along with precise sensing capabilities \cite{9705498}.
As a leading candidate technology for 6G, integrated sensing and communication (ISAC) systems integrate signal processing and hardware frameworks to enable wireless communication and environmental sensing in a single system. This integration is seen as a promising strategy for optimizing the use of available spectrum, energy resources, and hardware.

To efficiently integrate communication and sensing functions, the transceiver should operate in full-duplex (FD) mode, where the transmitter continuously receives the sensing echo signal at the receiver while transmitting the dual-function signal. However, a significant challenge posed by the FD mode is self-interference (SI) \cite{jasc_dongnan}. This issue is particularly problematic in FD ISAC systems, where the weaker sensing echo signals are often masked by direct signal interference between the transmit antennas and the receive antennas \cite{9363029}. Therefore, one of the key missions of ISAC systems is to tackle self-interference cancellation (SIC).
To address this, the authors in \cite{9443340} described radar processing of frequency-domain using a grid of fifth Generation(5G) new radio time-frequency resources or long-term evolution (LTE) and introduced efficient analog and digital SIC solutions designed for orthogonal frequency division multiplexing radar. The study in \cite{9838368} designed digital and analog transmit and receive beamformers, along with digital SIC units and active analog, aiming to maximize achievable rate of downlink and direction-of-arrival (DoA) accuracy while improving the performance of sensing including distance and detection target relative velocity estimation. The authors in \cite{Jasc_jiaohao} proposed the joint design of transmit and receive beamformers for the downlink users, a transmit precoder, and a receive combiner for the uplink users to solve the problem of missing sensing echoes caused by high residual SI.
The aforementioned studies rely on traditional SIC approaches in FD communication systems, which are divided into three steps: passive suppression, analog cancellation, and digital cancellation.  
However, traditional SIC approaches face challenges such as limited performance of physical isolation techniques in wideband systems due to receiver size and antenna configuration \cite{10258345}. Analog cancellation is hardware-intensive and costly, with residual SI and nonlinear distortion at high transmit power \cite{10198236}. Digital cancellation suffers from phase noise and is limited with analog techniques \cite{zongshu}. Therefore, exploring new SIC schemes for FD ISAC systems is essential.

Numerous existing studies have demonstrated the deployment of reconfigurable intelligent surface (RIS) can greatly improve the signal propagation environment in wireless communication systems \cite{RIS,9839223,10289894,10289881,10464418,WUqingqing}. RISs are planar arrays consisting of a significant quantity of passive reflective elements that can effectively enhance wireless signal coverage and improve the quality of communication by altering the angle of reflection and phase of the electromagnetic waves at much lower hardware and energy costs than conventional active antenna arrays \cite{RIS}.

Furthermore, RIS deployment in FD systems has a significant effect on reducing SI by using RIS to generate an appropriate offset signal in the analog domain to aid in SIC. This solution was first proposed by \cite{9839223}, which developed a greedy heuristic algorithm to design the optimal RIS phase, successfully reducing the leakage signal by an additional 59 dB in a narrowband environment. Consequently, several studies have been carried out to explore the possibility of deploying RIS in FD ISAC systems to reduce SI. For instance, the authors in \cite{10289894} and \cite{10289881} combined RIS in the near field of MIMO to integrate communication and sensing and suppress SI. Similarly, in the work of \cite{10464418}, researchers focused on how SI can be suppressed in monostatic ISAC scenarios by effectively applying RIS and further explored the interaction between SI and hardware limitations (e.g., low noise amplifier receivers). Additionally, the authors in \cite{WUqingqing} proposed an iterative algorithm co-designed for beamforming, signal processing, and power allocation in RIS-enabled FD ISAC systems.
However, the above studies still require the assistance of traditional SIC approaches or assume high-precision transmit beamforming. This is because RIS-assisted FD ISAC systems lack the ability to enhance both communication and sensing performance while efficiently achieving SIC (reduced to below noise power). Specifically, conventional RIS restricts both the transmitter and receiver to the same side, preventing the receiver from being positioned behind the RIS. To generate a sufficiently strong offset signal, the RIS needs to be placed very close to the FD transceiver (e.g., within 1 m) \cite{9839223}. In such proximity, the RIS can obstruct the signal propagation path, resulting in a reduced signal coverage area due to its half-space coverage characteristics.

  In recent years, Researchers developed simultaneous transmitting
and reflecting RIS (STAR-RIS)  \cite{9690478}, which not only enables 360-degree coverage of reflection and refraction \cite{10133841}, but also be able to reduce the SI to achieve FD \cite{Fang}. Specifically, the feature of STAR-RIS that allows the transmitter and receiver to be located on different sides enables STAR-RIS to be deployed in close proximity to the FD transceiver, which makes it possible not only to provide additional beamforming to the ISAC signals but also to provide a sufficiently large offset signal to reduce SI to the level below the noise power in the receive antennas.
The author in \cite{Fang} was the first to explore the effectiveness of Intelligent Omnidirectional Surfaces (IOS, equivalent to STAR-RIS) applied for SIC in an FD communication system, and the results showed that this deployment is effective in reducing SI without the traditional SIC approaches. Then, the study of \cite{wen} applied STAR-RIS to an FD secure communication system, which not only reduced the SI to the same level as traditional SIC approaches but also controlled the interference power received by the eavesdropper, thus significantly enhancing the confidentiality of the communication. 
However, research on STAR-RIS in ISAC is still at an early stage, and to the best of our knowledge, how to deploy STAR-RIS in FD ISAC systems to enhance the performance of communication and sensing, and the simultaneous cancellation of SI has not yet been explored.

In this paper, we propose a novel STAR-RIS-enabled FD ISAC system, featuring the deployment of STAR-RIS in close proximity to the FD BS. The main contributions of this paper are outlined as follows:

\begin{enumerate}
\item We are the first to deploy STAR-RIS to an FD ISAC system to not only enhance simultaneous communication and target sensing but also effectively reduce SI. In the proposed system, the FD BS transmits the ISAC signal through the STAR-RIS to both users and the detection target. The sensing echo returns through the STAR-RIS to the receive antennas. To reduce SI, the transmit signals are also reflected by the STAR-RIS to the receive antennas to couple the SI from the direct path.

\item We optimize the maximization of sensing signal-to-interference-plus-noise-ratio (SINR) and communication sum rate, which are crucial for improving sensing accuracy and overall communication performance, by jointly designing the beamforming at the FD BS, and the reflecting and refracting coefficients of STAR-RIS. These optimizations are constrained by the FD BS's maximum transmit power, the STAR-RIS reflection and refraction coefficient limitations, and the minimum SINR requirement for the alternate function for communication or sensing.

\item To address the non-convex optimization problems with multiple coupled parameters, we develop alternating optimization algorithms. Specifically, a semi-definite relaxation (SDR) algorithm is used for transmit beamforming vector design. For reflecting and refracting coefficients, we adopt the successive convex approximation (SCA) method and an SDR-based algorithm to handle the quartic and quadratic constraints.

\item Simulation results validate the effectiveness of the proposed algorithms that it can significantly improve both the sensing and communication performance compared to the benchmark using the traditional SIC approach.
\end{enumerate}

The remainder of this paper is structured as detailed below. The proposed model of the STAR-RIS-enabled FD ISAC system is outlined in Section
II, followed by the
problem formulation of the sensing SINR maximization and sum rate maximization.
Section III and Section IV illustrate the particulars of the proposed alternating algorithms of optimizing the beamformer of the transmitter, and the reflecting and refracting coefficients of the STAR-RIS, respectively. 
Section V showcases numerical outcomes to assess the efficacy of the proposed algorithm compared with a benchmark scheme. Our conclusion is presented in Section VI.

\textit{Notations:} Vectors are denoted by boldface lower-case letters, while matrices are denoted by boldface upper-case letters. The superscripts $(\cdot)^T$ and $(\cdot)^H$ denote the transpose and Hermitian transpose, respectively. The spaces $\mathbb{C}^{m \times n}$ and $\mathbb{R}^{m \times n}$ represent the complex and real matrix spaces of dimensions $m \times n$, respectively. The notations $\text{Tr}(\cdot)$, $\text{Rank}(\cdot)$, $\text{diag}(\cdot)$, and $[\cdot]_{i,j}$ are used to denote the trace, rank, diagonal matrix, and the $(i,j)$th entry of a matrix, respectively. The Euclidean norm of a vector is denoted by $\|\cdot\|$, and the absolute value of a scalar is denoted by $|\cdot|$. The expectation operation is represented by $\mathbb{E}\{\cdot\}$, the imaginary part of a complex number by $\Im$, and the real part by $\Re$. Additionally, $\odot$ and $\otimes$ represent the Hadamard product and the Kronecker product, respectively, and $\text{vec}(\cdot)$ denotes the vectorization operation. The element $(i,j)$ of matrix $\mathbf{W}$ is also denoted by $\mathbf{W}(m,m)$. The notation $\mathbf{W} \succeq 0$ indicates that $\mathbf{W}$ is positive semidefinite. Lastly, $\mathcal{O}(\cdot)$ is used to denote the big-O computational complexity notation.

\section{System Model and Problem Formulation}
In this section, we first describe the STAR-RIS-enabled FD-ISAC system. Then, the optimization problems of maximizing the SINR and the communication sum rate are formulated.
\subsection{System model}
\begin{figure}[!t]
        \centering
        \includegraphics*[width=90mm]{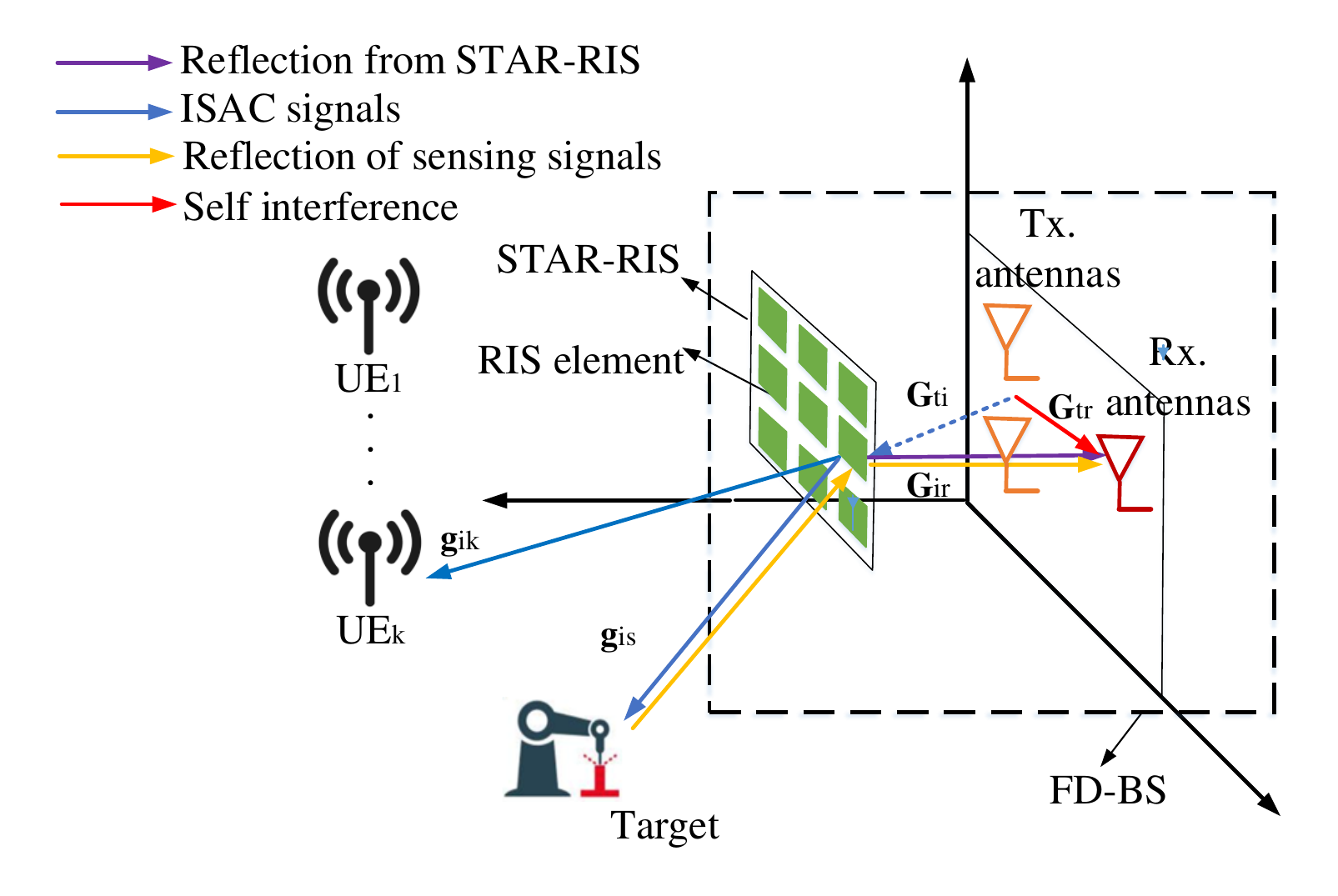}
       \caption{An illustration of STAR-RIS-enabled FD-ISAC system.}
        \label{fig:RIS-ISAC}
\end{figure}

Consider an FD-ISAC system with the assistance of a dual-side STAR-RIS, where an FD BS transmits integrated signals to send data to multiple users and to measure one detection target, while simultaneously receiving the echo signals of previous sensing. As depicted in Fig. 1, an $L$-element STAR-RIS is integrated into the BS to assist the transmission of ISAC signals and simultaneously reduce the SI in the BS's receive antennas.
Assuming that the users are equipped with a single antenna and the FD BS is equipped with $M$ transmit antennas and $N$ receive antennas. The STAR-RIS operates in the ES mode \cite{9690478}, ie., 
the signal energy incident on each element is partitioned between the transmitted and reflected signals, to allow each element to reflect and refract signals simultaneously.

To characterize the impact of STAR-RIS on the incident signals, the reflecting and refracting coefficients for the incidence from FD BS side are represented as ${\mathbf{\Phi}_b}={\text{diag}}\left\{ {{u_{b,1}}{e^{j{\mu _{b,1}}}},{u_{b,2}}{e^{j{\mu _{b,2}}}}, \ldots ,{u_{b,L}}{e^{j{\mu _{b,L}}}}} \right\} \in {\mathbb{C}^{L \times L}}$ and ${{\mathbf{\Theta}}_b}={\text{diag}}\left\{ {{v_{b,1}}{e^{j{\nu _{b,1}}}},{v_{b,2}}{e^{j{\nu _{b,2}}}}, \ldots ,{v_{b,L}}{e^{j{\nu _{b,L}}}}} \right\}\in{\mathbb{C}^{L \times L}}$, respectively, and $u_{b,l}$ and $v_{b,l}$ represent the reflecting and refracting amplitudes of the $l$th element, respectively, and $\mu_{b,l}$ and $\nu_{b,l}$ indicate the phase shifts for reflection and refraction at the $l$th element, respectively. Similarly, corresponding matrices for the incidence from user side are represented as ${{\mathbf{\Phi}}_u}={\text{diag}}\left\{ {{u_{u,1}}{e^{j{\mu _{u,1}}}},{u_{u,2}}{e^{j{\mu _{u,2}}}}, \ldots ,{u_{u,L}}{e^{j{\mu _{u,L}}}}} \right\} \in {\mathbb{C}^{L \times L}}$ and ${{\mathbf{\Theta}}_u}={\text{diag}}\left\{ {{v_{u,1}}{e^{j{\nu _{u,1}}}},{v_{u,2}}{e^{j{\nu _{u,2}}}}, \ldots ,{v_{u,L}}{e^{j{\nu _{u,L}}}}} \right\} \in {\mathbb{C}^{L \times L}}$, respectively.

Based on \cite{Dual-STAR}, the reflection and refraction coefficients are identical on both sides, thus $\mathbf\Phi_{b} = \mathbf\Phi_{u}$ and $\mathbf\Theta_{b} = \mathbf\Theta_{u}$. Therefore, only $\mathbf\Phi_{b}$ and $\mathbf\Theta_{b}$ are used for simplicity. According to \cite{STAR-2}, the constraints can be defined as follows:
\begin{subequations}\label{constraint_1}
\begin{equation}\label{1a}
u_{b,l}^2 + v_{b,l}^2 \leq 1,
\end{equation}
\begin{equation}\label{1b}
0 \leq u_{b,l}, v_{b,l} \leq 1,
\end{equation}
\begin{equation}\label{1c}
0 \leq \mu_{b,l}, \nu_{b,l} < 2\pi, \quad \forall l,
\end{equation}
\end{subequations}
\noindent where \eqref{1a} and \eqref{1b} represent the constraints of the total power of refraction and reflection, and constraints of the amplitude, respectively.

The ISAC signals transmitted from the FD BS are represented as
\begin{equation}
\mathbf{x} = \sum_{k=1}^{K} \mathbf{w}_k \mathbf{s}_k + \mathbf{w}_s \mathbf{s}_s,
\end{equation}
where $\mathbf{w}_k \in \mathbb{C}^{M \times 1}$ is the vector of downlink transmit beamforming for $K$ users, and $\mathbf{s}_k \in \mathbb{C}$ is the signal vector for the $k$th user. On the other hand, $\mathbf{w}_s \in \mathbb{C}^{M \times 1}$ and $\mathbf{s}_s \in \mathbb{C}$ are the beamforming vector and signal vector for detection target, respectively.

Then, the received signals of the $k$th user can be expressed as
\vspace{0.02em}
\begin{equation}\label{csignal_k}
{{y}_{k}} = {\mathbf{g}}_{i,k}^H{{\mathbf{\Theta}}_b}{{\mathbf{G}}_{ti}}{\mathbf{x}} + {n}_k,
\end{equation}
\noindent where ${{\mathbf{g}}_{i,k}} \in {\mathbb{C}^{L \times 1}}$ is the vector of channel coefficients between the STAR-RIS and the $k$th user, and ${{\mathbf{G}}_{ti}} \in {\mathbb{C}^{L \times M}}$ is the matrix of channel coefficients between the transmit antennas and the STAR-RIS, and ${{n}_k} \sim \mathcal{CN}\left( {0,\sigma _k^2} \right)$ represents the additive white Gaussian noise (AWGN) with zero mean and variances $\sigma_k^2$ of the $k$th user.

On the other hand, the reflected sensing signals received in the FD BS\footnote {
Assuming perfect CSI for all channels, including those with the target, is available for optimization. Details regarding channel estimation involving the STAR-RIS are outside the scope of this paper and can be found in \cite{CSI1,CSI2}. The proposed algorithm, leveraging perfect CSI, establishes an upper bound on secrecy performance. Additionally, in Section V, we provide results using imperfect CSI for comparative analysis.} can be expressed by
\vspace{0.02em}
\begin{align}\label{sensingsignal}
    {{\mathbf{y}}_{s}} =&  \ \beta_{s}{{\mathbf{G}}_{ir}^H}{\mathbf{\Theta}}_{b}^H{\mathbf{g}}_{is}^H{\mathbf{g}}_{is}{\mathbf{\Theta}}_{b}{{\mathbf{G}}_{ti}}{\mathbf{x}} +   \left( {\mathbf{G}}_{tr} + {\mathbf{G}}_{ir}^H{{\mathbf{\Phi}}_b}{{\mathbf{G}}_{ti}} \right) {\mathbf{x}} \nonumber  \\
    &+ \sum\limits_{q = 1}^Q \beta_{q}{{\mathbf{G}}_{ir}^H}{\mathbf{\Theta}}_{b}^H{\mathbf{g}}_{i,q}^H{\mathbf{g}}_{i,q}{\mathbf{\Theta}}_{b}{{\mathbf{G}}_{ti}}{\mathbf{x}} + {{\mathbf{n}}_s},
\end{align}
 \noindent where the first term denotes the expected echo signals from the detection target, the second term expresses the signals of SI through both the direct path and reflected path from the STAR-RIS, and the third term is the combined signals from $Q$ undesired sources whose echo signals cause interference for sensing. 
$\beta_{s}$ and $\beta_{q}$ are the reflection factors on the target and the $q$th interference source, respectively. ${\mathbf{g}}_{is} \in {\mathbb{C}^{L \times 1}}$  and ${\mathbf{g}}_{i,q} \in {\mathbb{C}^{L \times 1}}$ are the channel coefficients between the STAR-RIS and the detection target, and those between the STAR-RIS and the $q$th interference source, respectively. ${{\mathbf{G}}_{tr}} \in {\mathbb{C}^{N \times M}}$ and ${{\mathbf{G}}_{ir}} \in {\mathbb{C}^{L \times N}}$ are the matrices, 
the elements of which denote the coefficients of the channels between the transmit and the receive antennas of the FD BS, as well as between STAR-RIS elements and the receive antennas of the FD BS, respectively. ${\mathbf{n}_s} \sim \mathcal{CN}\left( 0,\sigma _s^2 \right)$ represents AWGN with  zero mean and variances $\sigma_s^2$ at the receive antennas of the FD BS.  

Assuming the channel between the transmit antennas of the FD BS and the STAR-RIS, and the channel between the receive antennas of the FD BS and the STAR-RIS as near-field line-of-sight (LOS) channels, considering the small distance $r$ between the STAR-RIS and the antenna panel. 
To determine the related coefficients of channels, we define the relative position between the $m$th transmit antenna and the $l$th element of STAR-RIS by $\left( {{r_{m,l}},\theta _{m,l},\phi _{m,l}} \right)$. Taking the location of the $l$th element as the reference origin, thereby the relative distance ${r_{m,l}} \geq 0$, the elevation angle $\theta _{m,l} \in \left[ {0,\frac{\pi }{2}} \right]$ and the azimuth angle $\phi _{m,l} \in \left[ {0,2\pi } \right]$ can be defined, respectively. 
Similarly, the relative positions between the $n$th receive antenna at the FD BS and the $l$th STAR-RIS element are denoted by $\left( {r_{n,l}, \theta_{n,l}, \phi_{n,l}} \right)$, while those between the $m$th transmit antenna and the $n$th receive antenna of the FD BS are represented as $\left( {r_{n,m}, \theta_{n,m}, \phi_{n,m}} \right)$. With the above definitions, the corresponding channel coefficients can derived as
\begin{equation}
\begin{gathered}
  {{\mathbf{G}}_{tr}} = \left[ {\frac{{\lambda \sqrt {{G^t}\left( {{\theta _{n,m}},{\phi _{n,m}}} \right){G^r}\left( {{\theta _{n,m}},{\phi _{n,m}}} \right)} }}{{4\pi r_{n,m}^{{\kappa  \mathord{\left/
 {\vphantom {\kappa  2}} \right.
 \kern-\nulldelimiterspace} 2}}}}} \right. \hfill \\
  {\left. { \qquad\quad \times \left( {\sqrt {\frac{\vartheta}{{\vartheta + 1}}} {e^{ - j\frac{{2\pi {r_{n,m}}}}{\lambda }}} \!+\! \sqrt {\frac{1}{{\vartheta + 1}}} g_{tr}^{nlos}} \right)} \right]_{n,m}}, \hfill \\ 
\end{gathered} 
\end{equation}

\begin{align}
&{{\mathbf{G}}_{ir}} = {\left[ {\frac{{\lambda \sqrt {{G^r}\left( {{\theta _{n,l}},{\phi _{n,l}}} \right)} }}{{4\pi {r_{n,l}}}}{e^{ - j\frac{{2\pi {r_{n,l}}}}{\lambda }}}} \right]_{n,l}}, \\
&{{\mathbf{G}}_{ti}} = {\left[ {\frac{{\lambda \sqrt {{G^t}\left( {{\theta _{m,l}},{\phi _{m,l}}} \right)} }}{{4\pi {r_{m,l}}}}{e^{ - j\frac{{2\pi {r_{m,l}}}}{\lambda }}}} \right]_{m,l}},
\end{align}

\noindent where $\lambda $ denotes the wavelength, ${G^t}\left( {\theta,\phi } \right)$ and ${G^r}\left( {\theta,\phi } \right)$ represent the gains of FD BS's transmit and receive antennas, respectively.
Additionally, $\vartheta$ is the Rician factor, and $\kappa$ denotes the pathloss exponent. and $g_{tr}^{{nlos}}$ represents the non-line-of-sight (NLoS) channel component, which is generalized using zero-mean and unit-variance circularly symmetric complex Gaussian random variables.

On the other hand, the channels between STAR-RIS elements and the users and the target are considered as far-field LOS channels, thus the related channel coefficients can be expressed as follows \cite{8627376}:
\vspace{0.02em}
\begin{equation}
{{\mathbf{g}}_{i,k}} \!=\!  \frac{\sqrt {g_0} }{r_{i,k}} {\mathbf{a}_{i,k}},
\end{equation}
\vspace{0.02em}
\begin{equation}
{{\mathbf{g}}_{is}} \!=\!  \frac{\sqrt {g_0} }{r_{is}} {\mathbf{a}_{is}},
\end{equation}

\noindent where $g_0$ represents the channel gain in the reference distance $1$~m, $r_{i,k}$ and $r_{i,k}$ are the distances from the origin point of the STAR-RIS to the $k$th user, and from the origin point of the STAR-RIS to the detection target, respectively. Furthermore, ${\mathbf{a}_{i,k}}$ and ${\mathbf{a}_{is}}$ are the steering vector of STAR-RIS to the $k$th user and to the detection target, which can be calculated as follows: 
\vspace{0.02em}
\begin{align}
 {{\mathbf{a}}_{i,{\text{z}}}} = &\left( {1, \ldots ,{e^{ - j\frac{{2\pi d}}{\lambda }\sin {\theta _{i,{\text{z}}}}\left( {{L_x} - 1} \right)\cos {\varphi _{i,{\text{z}}}}}}} \right) ^T\nonumber \\ &\otimes \left( {1, \ldots ,{e^{ - j\frac{{2\pi d}}{\lambda }\sin {\theta _{i,{\text{z}}}}\left( {{L_y} - 1} \right)\sin {\varphi _{i,{\text{z}}}}}}} \right)^T,   
\end{align}
where ${\text{z}} \in \left\{ {1,\ldots,k,s} \right\}$, $d$ is the spacing between two adjacent elements in the same line or column, $\theta_{i,\text{z}}$ and $\phi_{i,\text{z}}$ are the corresponding angels, respectively. $L_x$ and $L_y$ are the numbers of elements in each line and column at the squared STAR-RIS, respectively,  leading to the total number of elements given by $L=L_x \times L_y$.  

Thus, the SINR of the received communication signals in the user $k$ can be represented as follows:
\begin{equation}
\begin{aligned}
{\gamma_{k}} =&  \frac{\mathbb{E} \{|{\mathbf{g}}_{i,k}^H{{\mathbf{\Theta}}_b}{{\mathbf{G}}_{ti}}\mathbf{w}_{k}\mathbf{s}_k|^2\}}{\sum\limits_{j \neq k}^K \mathbb{E}\{|{\mathbf{g}}_{i,k}^H{{\mathbf{\Theta}}_b}{{\mathbf{G}}_{ti}}\mathbf{w}_{j}\mathbf{s}_j |^2\} + \mathbb{E}\{|{\mathbf{g}}_{i,k}^H{{\mathbf{\Theta}}_b}{{\mathbf{G}}_{ti}}\mathbf{w}_{s}\mathbf{s}_s|^2\} + {\sigma _k^2} } \\ 
=& \frac{ |{\mathbf{g}}_{i,k}^H{{\mathbf{\Theta}}_b}{{\mathbf{G}}_{ti}}\mathbf{w}_{k}|^2}{\sum\limits_{j \neq k}^K |{\mathbf{g}}_{i,k}^H{{\mathbf{\Theta}}_b}{{\mathbf{G}}_{ti}}\mathbf{w}_{j} |^2 + |{\mathbf{g}}_{i,k}^H{{\mathbf{\Theta}}_b}{{\mathbf{G}}_{ti}}\mathbf{w}_{s}|^2 + {\sigma _k^2} } \label{SINR_K}.
\end{aligned}
\end{equation}

Furthermore, the SINR of the received echo signals of sensing is given as
\begin{equation}
\begin{aligned}
{\gamma_{s}} =&  \frac{\mathbb{E} \{|\mathbf{P}_s\mathbf{x}|^2\}}{\mathbb{E}\{|{{\mathbf{G}}_{SI}}\mathbf{x} |^2\} +\sum\limits_{q = 1}^Q \mathbb{E}\{|\mathbf{P}_q{\mathbf{x}}|^2\} + {\sigma _s^2} } \\ 
=& \frac{ | \beta_{s}{{\mathbf{G}}_{ir}^H}{\mathbf{\Theta}}_{b}^H{\mathbf{g}}_{is}{\mathbf{g}}_{is}^H{\mathbf{\Theta}}_{b}{{\mathbf{G}}_{ti}}\mathbf{w}|^2}
{ |\mathbf{G}_{SI}\mathbf{w}|^2 +
\sum\limits_{q = 1}^Q|\beta_{q}{{\mathbf{G}}_{ir}^H}{\mathbf{\Theta}}_{b}^H{\mathbf{g}}_{i,q}{\mathbf{g}}_{i,q}^H{\mathbf{\Theta}}_{b}{\mathbf{G}}_{ti}\mathbf{w}|^2 + {\sigma _s^2} } \label{SINR_t},
\end{aligned}
\end{equation}

\noindent where 
$\mathbf{w}=[ \mathbf w_1,..., \mathbf w_K, \mathbf w_s] \in {\mathbb{C}^{M \times (K+1)}}$, and
$\mathbf{G}_{SI} = {\mathbf{G}}_{tr} + {\mathbf{G}}_{ir}^H{{\mathbf{\Phi}}_b}{{\mathbf{G}}_{ti}}$ denotes the channel coefficients matrix for SI. 
\subsection{Problem formulation}
The objective of this paper is to improve the sensing accuracy and overall communication performance of the proposed system. To achieve this, two optimization problems are designed. For the first optimization problem, we jointly optimize the transmit beamforming vector at the FD BS and the STAR-RIS coefficients to maximize the sensing SINR, while guaranteeing the minimum communication SINR, the maximum transmit power at the FD BS, and the reflection and refraction coefficients constraints of the STAR-RIS. The corresponding problem is formulated as

\subsubsection{P1 Maximization of sensing SINR}
\begin{small}\begin{align}
\mathop {\max }\limits_{\mathbf{w}_k,\mathbf{w}_s,\mathbf{\Phi}_b,\mathbf{\Theta}_b}&\ \ {\gamma_{s}}  \label{P1_OF}\\
{\text{s.t.}}\ \ & \ \   {\text{Tr}}\left(\sum _{k=1}^{K}  {{\mathbf{w}_k}{\mathbf{w}_k^H}} + {\mathbf{w}_s}{\mathbf{w}_s^H} \right) \leq P_t, \tag{\ref{P1_OF}{a}}  \label{P1a} \\
& \ \  \gamma_k  \geq \gamma_{req}, \forall k,  \label{limit_r} \tag{\ref{P1_OF}{b}}   \\
& \ \  \eqref{1a}-\eqref{1c}, \tag{\ref{P1_OF}{c}}  \label{P1c}
\end{align}\end{small}

 \noindent where \eqref{P1a} denotes the power constraint of the beamforming with a maximum power $P_t$, $\text{(13b)}$
 guarantees the minimum communication SINR with $\gamma_{req}$ indicating the required SINR for communication signals, and  $\text{(13c)}$ denotes the reflection and refraction coefficient constraints of STAR-RIS. 

Regarding the second optimization problem, we jointly optimize the transmit beamforming vector at the FD BS and the STAR-RIS coefficients to maximize the total sum rate for downlink communication, while guaranteeing the minimum communication SINR, the maximum transmit power at the FD BS, and the reflection and refraction coefficients constraints of the STAR-RIS. The corresponding problem is formulated as

\subsubsection{P2 Maximization of the total sum rate}
\begin{small}\begin{align}
\mathop {\max }\limits_{\mathbf{w}_k,\mathbf{w}_s,\mathbf{\Phi}_b,\mathbf{\Theta}_b}&\ \ \sum\limits_{k = 1}^K {\log_2 (1+\gamma_{k})}  \label{P2}\\
{\text{s.t.}}\ \ & \ \   \gamma_s  \geq \gamma_{min}, \tag{\ref{P2}{a}}   \\
& \ \  \eqref{P1a}, \eqref{P1c},    \tag{\ref{P2}{b}} \end{align}\end{small} 

\noindent where $\text{(14a)}$ guarantees the minimum communication SINR, with $\gamma_{min}$ indicating the required minimum SINR for the sensing. $\text{(14b)}$ denotes the maximum power constraint of the beamforming and the reflection and refraction coefficients constraints.

\section{Joint Beamforming and Reflecting Design for Sensing SINR Maximization}
We handle (13) in this section. We regard (13) as 
an optimization problem that is both non-convex and NP-hard due to the joint optimization of multiple coupled variables, making it challenging to solve directly. Therefore, we divide it into two sub-problems: optimizing the ISAC beamforming and optimizing the STAR-RIS coefficients.
\subsection{ Transmit Beamforming Optimization}
In this subsection, the sub-problem of transmit beamforming defined in \eqref{P1_OF} is examined, we consider that the STAR-RIS coefficients $\mathbf{\Phi}_b$ and $\mathbf{\Theta}_b$ are given. 
We start with rewriting the SINRs for convenience. \eqref{SINR_K} can be reformulated as
\begin{equation}
\begin{aligned}
{\gamma_{k}} =\frac{\text{Tr}\left({\mathbf{G}}_{t,k}\mathbf{w}_{k}\mathbf{w}_{k}^H{{\mathbf{G}}_{t,k}^H}\right)}{{\sum\limits_{j \neq k}^K\text{Tr}{\left({{\mathbf{G}}_{t,k}\mathbf{w}_{j}\mathbf{w}_{j}^H{{\mathbf{G}}_{t,k}^H}}\right)}} +\text{Tr}\left({{\mathbf{G}}_{t,k}\mathbf{w}_{s}\mathbf{w}_{s}^H{{\mathbf{G}}_{t,k}^H}}\right) + {\sigma _k^2}},
\end{aligned}
\end{equation}
where ${\mathbf{G}_{t,k}}$ = ${\mathbf{g}}_{i,k}^H{{\mathbf{\Theta}}_b}{{\mathbf{G}}_{ti}}$, and \eqref{SINR_t} can be reformulated as
\begin{equation}
\begin{aligned}
{\gamma_{s}} 
=\frac{\text{Tr}\left(\mathbf{P}_s\mathbf{W}\mathbf{P}_s^H\right)}{{\text{Tr}{\left({{\mathbf{G}}_{SI}\mathbf{W}{{\mathbf{G}}_{SI}^H}}\right)}} +\text{Tr}\left({{\mathbf{P}}_q\mathbf{W}{{\mathbf{P}}_q^H}}\right) + {\sigma _s^2}},
\end{aligned}
\end{equation}
where $\mathbf{P}_s   = \beta_{s}{{\mathbf{G}}_{ir}^H}{\mathbf{\Theta}}_{b}^H{\mathbf{g}}_{is}{\mathbf{g}}_{is}^H{\mathbf{\Theta}}_{b}{{\mathbf{G}}_{ti}},  
\mathbf{P}_q   = \beta_{q}{{\mathbf{G}}_{ir}^H}$${\mathbf{\Theta}}_{b}^H{\mathbf{g}}_{i,q}{\mathbf{g}}_{i,q}^H{\mathbf{\Theta}}_{b}{\mathbf{G}}_{ti}, q=1,2,...,Q,$ and\\ 
\begin{small}\begin{align} \mathbf W \triangleq \mathbb E \{ \mathbf x \mathbf x^{H}\} =\sum _{k=1}^{K} \mathbf w_{k} \mathbf w_{k}^{H} + \mathbf w_{s}\mathbf w_{s}^H
\end{align}\end{small}

\noindent represents the covariance matrix of the ISAC signal.

 Subsequently, omitting constant terms simplifies the transmit beamforming sub-problem to the following expression:
\vspace{0.02em}
\begin{small}\begin{align}
\mathop {\max }\limits_{\{\mathbf{w}_k\}_{k=1}^{K},\mathbf{w}_s}&\ \ {\gamma_{s}}  \label{P1_OF1}\\
{\text{s.t.}}\ \ & \ \   {\text{Tr}}\left( \sum _{k=1}^{K} {{\mathbf{w}_k}{\mathbf{w}_k^H}} + {\mathbf{w}_s}{\mathbf{w}_s^H} \right) \leq P_t, \tag{\ref{P1_OF1}{a}}  \label{P11a} \\
& \ \  \gamma_k  \geq \gamma_{req}, \forall k \label{limit_r} \tag{\ref{P1_OF1}{b}}.
\end{align}\end{small}

The above problem exhibits non-convexity owing to its complex quadratic fractional objective function and SINR constraints. The SDR
technique \cite{5447068} is applied to tackle this difficulty. Define ${\mathbf{W}_{k}} =\mathbf{w}_{k}\mathbf{w}_{k}^H,{\mathbf{W}_{j}} =\mathbf{w}_{j}\mathbf{w}_{j}^H$ and ${\mathbf{W}_{s}} =\mathbf{w}_{s}\mathbf{w}_{s}^H $. The problem \eqref{P1_OF1} can be reformulated as
\begin{small}\begin{align}
&\mathop {\max }\limits_{\{\mathbf{W}_k\}_{k=1}^{K} ,\mathbf{W}_s }  \ \  \frac{\text{Tr}\left(\mathbf{P}_s\mathbf{W}\mathbf{P}_s^H\right)}{{\text{Tr}{\left({{\mathbf{G}}_{SI}\mathbf{W}{{\mathbf{G}}_{SI}^H}}\right)}} +\text{Tr}\left({{\mathbf{P}}_q\mathbf{W}{{\mathbf{P}}_q^H}}\right) + {\sigma _s^2}}  \label{P1_OF2}\\
&{\text{s.t.}}\ \  \ \  
{\text{Tr}}\left( {{\sum _{k=1}^{K}\mathbf{W}_k}}+ {\mathbf{W}_s} \right) \leq P_t,\tag{\ref{P1_OF2}{a}}\\
 &  \frac{\text{Tr}\left({\mathbf{G}}_{t,k}\mathbf{W}_{k}{{\mathbf{G}}_{t,k}^H}\right)}{{\sum\limits_{j \neq k}^K\text{Tr}{\left({{\mathbf{G}}_{t,k}\mathbf{W}_{j}{{\mathbf{G}}_{t,k}^H}}\right)}} +\text{Tr}\left({{\mathbf{G}}_{t,k}\mathbf{W}_{s}{{\mathbf{G}}_{t,k}^H}}\right) + {\sigma _k^2}}\geq \gamma_{req},\forall k, \tag{\ref{P1_OF2}{b}}\\
 & \ \ \mathbf{W}_k \succeq 0, \forall k,    \hspace{1cm}   \mathbf{W}_s \succeq 0. \tag{\ref{P1_OF2}{c}}
\end{align}\end{small}

Note that the rank-one constraint has been omitted. However, problem \eqref{P1_OF2} remains non-convex due to its fractional objective function. Thus, we rewritten the objective function using the Dinkelbach method as
\vspace{0.02em}
\begin{equation}
\begin{small}\begin{aligned}
&\text{Tr}\left(\mathbf{P}_s\mathbf{W}\mathbf{P}_s^H\right) \\
&
- y^{(n)}
\left({\text{Tr}{\left({{\mathbf{G}}_{SI}\mathbf{W}{{\mathbf{G}}_{SI}^H}}\right)}} +\text{Tr}\left({{\mathbf{P}}_q\mathbf{W}{{\mathbf{P}}_q^H}}\right) + {\sigma _s^2}\right)
\end{aligned}\end{small} 
\end{equation}

\noindent where $\mathbf{W}$ is an affine function of $\{{\mathbf{\{{w}}_k}\}_{k=1}^{K},\{\mathbf{w}_s\}\}$, and $y^{(n)}$ denotes the value of $y$ at the $n$th iteration, which be defined as
\begin{small}\begin{align}
y^{(n)} = \left( \frac{\text{Tr}\left(\mathbf{P}_s\mathbf{W}\mathbf{P}_s^H\right)}{{\text{Tr}{\left({{\mathbf{G}}_{SI}\mathbf{W}{{\mathbf{G}}_{SI}^H}}\right)}} +\text{Tr}\left({{\mathbf{P}}_q\mathbf{W}{{\mathbf{P}}_q^H}}\right) + {\sigma _s^2}}\right)^{(n-1)}
\end{align}\end{small}

Therefore, the problem  \eqref{P1_OF2} can be expressed as 
\begin{small}\begin{align}
&\mathop {\max }\limits_{\{\mathbf{W}_k\}_{k=1}^{K},\mathbf{W}_s}\ \ \text{(20)}  \label{P1_OF3}\\
{\text{s.t.}}\ \ 
& \ \ \text{(19a), (19b), (19c)}, \tag{\ref{P1_OF3}{c}}
\end{align}\end{small}

\noindent which is convex and the corresponding optimal solutions can be obtained by convex optimization tools, such as CVX \cite{grant2009cvx}. Built upon the rank reduction theorem in Theorem 3.2 of \cite{rank1}, the problem \eqref{P1_OF3} has an optimal solution $\{\mathbf{\{{W}}^*_k\}_{k=1}^{K},\{\mathbf{W}^*_s\}\}$  satisfying the rank-one constraint. Thus, the optimal $\{{\mathbf{\{{w}}^*_k}\}_{k=1}^{K},\{\mathbf{w}^*_s\}\}$ can be recovered by eigenvalue decomposition.

\subsection{Optimization of the Coefficients for the STAR-RIS }
In this subsection, we optimize the STAR-RIS coefficients $\mathbf{\Phi}_b$ and $\mathbf{\Theta}_b$ when the beamforming matrices ${\mathbf{\{{w}}_k}\}_{k=1}^{K}$ and $\{\mathbf{w}_s\}$ are fixed. Specifically, the sub-problem can be expressed as
\begin{small}\begin{align}
\mathop {\max }\limits_{\mathbf{\Phi}_b,\mathbf{\Theta}_b}&\ \ {\gamma_{s}} \label{P2_OF}\\
{\text{s.t.}}
& \ \  \gamma_k  \geq \gamma_{req}, \forall k,    \tag{\ref{P2_OF}{a}} \label{23b}  \\
& \ \  \eqref{1a}-\eqref{1c}. \label{limit_r} \tag{\ref{P2_OF}{b}}   
\end{align}\end{small}
Since the constraints \eqref{23b} -- \eqref{limit_r}  and the objective function are not convex, new variables $\mathbf{v}_1=[{{v_{b,1}}{e^{j{\nu _{b,1}}}},{v_{b,2}}{e^{j{\nu _{b,2}}}}, \ldots ,{v_{b,L}}{e^{j{\nu _{b,L}}}}}]^H$ and  $\mathbf{v}_2 =[{{u_{b,1}}{e^{j{\mu _{b,1}}}},{u_{b,2}}{e^{j{\mu _{b,2}}}}, \ldots ,{u_{b,L}}{e^{j{\mu _{b,L}}}}}] ^T$ are introduced, define $
\overline{\mathbf v}_1 =\begin{bmatrix}
\mathbf v_1 \\
1
\end{bmatrix}$.

For solving the non-convex constraint in \eqref{23b}, we define $\mathbf{u}_{b,c} = \text{diag}(\mathbf{g}_{i,b}^H){{\mathbf{G}}_{ti}}\mathbf{w}_{c},$ where $ b=1,...,K, c=1,...,K,s $. Based on the principle of matrix operations, the term $|{\mathbf{g}}_{i,k}^H{{\mathbf{\Theta}}_b}{{\mathbf{G}}_{ti}}\mathbf{w}_{k}|^2$ can be rewritten as $|\mathbf{v}_1^H\mathbf{u}_{k,k}|^2$, which can be reformulated to $\text{Tr}(\overline{\mathbf v}_1^H\mathbf{U}_{k,k}\overline{\mathbf v}_1)$, where $ \mathbf U_{b,c} = 
\begin{bmatrix}
\mathbf u_{b,c} \mathbf u_{b,c}^H & 0 \\
0 & 0
\end{bmatrix}$. Thus, \eqref{23b} can be re-expressed as
\begin{equation}
\begin{aligned}
\frac{ \text{Tr}(\overline{\mathbf v}_1^H\mathbf{U}_{k,k}\overline{\mathbf v}_1)}{\sum\limits_{j \neq k}^K \text{Tr}(\overline{\mathbf v}_1^H\mathbf{U}_{k,j}\overline{\mathbf v}_1) + \text{Tr}(\overline{\mathbf v}_1^H\mathbf{U}_{k,s}\overline{\mathbf v}_1) + {\sigma _k^2} } \geq \gamma_{req}, \forall k.   \label{SINR_K1}
\end{aligned}
\end{equation}

 Define $\mathbf{V}_1 =\overline{\mathbf v}_1 \overline{\mathbf v}_1^H$, where $\mathbf{V}_1\succeq 0$ and $\text{rank}(\mathbf{V}_1)=1$. Due to the cyclic property of the trace operation, \eqref{SINR_K1} can be further transformed as
\begin{equation}
\begin{aligned}
\frac{ \text{Tr}(\mathbf{U}_{k,k}{\mathbf V}_1)}{\sum\limits_{j \neq k}^K \text{Tr}(\mathbf{U}_{k,j}{\mathbf V}_1) + \text{Tr}(\mathbf{U}_{k,s}{\mathbf V}_1) + {\sigma _k^2} } \geq \gamma_{req}, \forall k.  \label{SINR_K2}
\end{aligned}
\end{equation}

For the non-convex term in the objective function, it is a fractional programming (FP) formulation and the terms are quartic functions with respect to the variables $\mathbf{\Phi}_b$ and $\mathbf{\Theta}_b$, which makes it challenging to solve. Firstly, presenting the objective function with $\mathbf{A},\mathbf{B} $ and $\mathbf{C}$.
\begin{equation}
\begin{aligned}
\frac{\mathbf A}{\mathbf B +\mathbf C + {\sigma _s^2}} \label{bianhua1},
\end{aligned}
\end{equation}
where
\begin{equation}
\begin{aligned}
 \mathbf A  = | \beta_{s}{{\mathbf{G}}_{ir}^H}{\mathbf{\Theta}}_{b}^H{\mathbf{g}}_{is}{\mathbf{g}}_{is}^H{\mathbf{\Theta}}_{b}{{\mathbf{G}}_{ti}}\mathbf{w}|^2   \label{A},
\end{aligned}
\end{equation}
\begin{equation}
\begin{aligned}
 \mathbf B =|\mathbf{G}_{SI}\mathbf{w}|^2 \label{B},
\end{aligned}
\end{equation}
\begin{equation}
\begin{aligned}
\mathbf C  = \sum\limits_{q = 1}^Q|\beta_{q}{{\mathbf{G}}_{ir}^H}{\mathbf{\Theta}}_{b}^H{\mathbf{g}}_{i,q}{\mathbf{g}}_{i,q}^H{\mathbf{\Theta}}_{b}{\mathbf{G}}_{ti}\mathbf{w}|^2  \label{C}.
\end{aligned}
\end{equation}

Note that the above expressions \eqref{bianhua1}, \eqref{A}, \eqref{B}, and \eqref{C} are still non-convex. Starting with \eqref{B}, the term  $|\mathbf{G}_{SI}\mathbf{w}|^2$ can be expanded as 
\begin{equation}
\begin{aligned}
&\text{Tr}(({\mathbf{G}}_{tr} + {\mathbf{G}}_{ir}^H{{\mathbf{\Phi}}_b}{{\mathbf{G}}_{ti}})\mathbf w \mathbf w^H({{\mathbf{G}}_{ti}^H}{{\mathbf{\Phi}}_b^H}\mathbf{G}_{ir}+{\mathbf{G}}_{tr}^H))\\
=&\text{Tr}({\mathbf{G}}_{tr}\mathbf w \mathbf w^H{{\mathbf{G}}_{ti}^H}{{\mathbf{\Phi}}_b^H}\mathbf{G}_{ir}) + 
\text{Tr}( {\mathbf{G}}_{ir}^H{{\mathbf{\Phi}}_b}{{\mathbf{G}}_{ti}}\mathbf w \mathbf w^H{{\mathbf{G}}_{ti}^H}{{\mathbf{\Phi}}_b^H}\\&\mathbf{G}_{ir})+
\text{Tr}({\mathbf{G}}_{ir}^H{{\mathbf{\Phi}}_b}{{\mathbf{G}}_{ti}})\mathbf w \mathbf w^H{\mathbf{G}}_{tr}^H)
+\text{Tr}({\mathbf{G}}_{tr}\mathbf w \mathbf w^H{\mathbf{G}}_{tr}^H)\label{27}.
\end{aligned}
\end{equation}

We can simplify \eqref{27} as
\begin{equation}
\begin{aligned}
\text{Tr}({{\mathbf{\Phi}}_b^H}\mathbf{X})+\text{Tr}({{\mathbf{\Phi}}_b^H}\mathbf{Y}{\mathbf{\Phi}}_b\mathbf{Z})+
\text{Tr}({{\mathbf{\Phi}}_b}\mathbf{X}^H)+\mathbf{D},
\end{aligned}
\end{equation}
where $\mathbf{X}=\mathbf{G}_{ir}{\mathbf{G}}_{tr}\mathbf w \mathbf w^H{{\mathbf{G}}_{ti}^H}$, $\mathbf{Y}={\mathbf{G}}_{ir}{\mathbf{G}}_{ir}^H$, $\mathbf{Z}={{\mathbf{G}}_{ti}}\mathbf w \mathbf w^H{{\mathbf{G}}_{ti}^H}$ and $\mathbf{D}=\text{Tr}({\mathbf{G}}_{tr}\mathbf w \mathbf w^H{\mathbf{G}}_{tr}^H)$.
By using the matrix operation principle in [\citealp{zhang2017matrix}, (1.10.6)], the above expression can be rewritten as
\begin{equation}
\begin{aligned}
&\mathbf{v}_2^T\text{diag}\{\mathbf{X}\} + \mathbf{v}_2^H (\mathbf{Y}\odot\mathbf{Z})\mathbf{v}_2+\text{diag}\{\mathbf{X}^H\}\mathbf{v}_2^{*}+\mathbf{D}\\
=&\mathbf{v}_2^H (\mathbf{Y}\odot\mathbf{Z})\mathbf{v}_2 + 2\Re\{\mathbf{v}_2^T\text{diag}\{\mathbf{X}\}\}+\mathbf{D}.
\end{aligned}
\end{equation}

Thus, \eqref{B} can be reformulated as
\begin{equation}
\begin{aligned}
 \mathbf B = \mathbf{v}_2^H (\mathbf{Y}\odot\mathbf{Z})\mathbf{v}_2 + 2\Re\{\mathbf{v}_2^T\text{diag}\{\mathbf{X}\}\}+\mathbf{D}, \label{SI}
\end{aligned}
\end{equation}
which is convex with respect to $\mathbf{v}_2$.

Then, for \eqref{A}, based on the principle of matrix operations, the term $| \beta_{s}{{\mathbf{G}}_{ir}^H}{\mathbf{\Theta}}_{b}^H{\mathbf{g}}_{is}{\mathbf{g}}_{is}^H{\mathbf{\Theta}}_{b}{{\mathbf{G}}_{ti}}\mathbf{w}|^2$ can be reformulated as
\begin{equation}
\begin{aligned}
&\text{Tr}((\beta_{s})^2{\mathbf{G}}_{ir}^H\text{diag}({\mathbf{g}}_{is})\mathbf{v}_1\mathbf{v}_1^H\text{diag}({\mathbf{g}}_{is}^H){{\mathbf{G}}_{ti}}\mathbf{w}\mathbf{w}^H{{\mathbf{G}}_{ti}^H}\text{diag}({\mathbf{g}}_{is})\\
&\mathbf{v}_1\mathbf{v}_1^H\text{diag}({\mathbf{g}}_{is}^H){\mathbf{G}}_{ir}) =
\text{Tr}((\beta_{s})^2\text{diag}({\mathbf{g}}_{is}^H){\mathbf{G}}_{ir}{\mathbf{G}}_{ir}^H\text{diag}({\mathbf{g}}_{is})\\&
\mathbf{v}_1\mathbf{v}_1^H\text{diag}({\mathbf{g}}_{is}^H){{\mathbf{G}}_{ti}}\mathbf{w}\mathbf{w}^H{{\mathbf{G}}_{ti}^H}\text{diag}({\mathbf{g}}_{is})\mathbf{v}_1\mathbf{v}_1^H) .\end{aligned}\label{354}
\end{equation}

Since $\text{Tr}(\mathbf{e}\mathbf{f}\mathbf{g}\mathbf{h})=(\text{vec}(\mathbf{h}^T))^T(\mathbf{g}^T\otimes\mathbf{e})\text{vec}(\mathbf{f})$, where $\otimes$ denotes
the Kronecker product, \eqref{354} can be expressed as 
\begin{equation}
\begin{aligned}
&\text{Tr}((\beta_{s})^2\text{diag}({\mathbf{g}}_{is}^H){\mathbf{G}}_{ir}{\mathbf{G}}_{ir}^H\text{diag}({\mathbf{g}}_{is})
\mathbf{v}_1\mathbf{v}_1^H\text{diag}({\mathbf{g}}_{is}^H)\\&{{\mathbf{G}}_{ti}}\mathbf{w}\mathbf{w}^H{{\mathbf{G}}_{ti}^H}\text{diag}({\mathbf{g}}_{is})\mathbf{v}_1\mathbf{v}_1^H) = 
\hat{\mathbf{v}}_1^H\mathbf{P}\hat{\mathbf{v}}_1\label{30},
\end{aligned}
\end{equation}
where $\mathbf{P} =(\text{diag}({\mathbf{g}}_{is}^H){{\mathbf{G}}_{ti}}\mathbf{w}\mathbf{w}^H{{\mathbf{G}}_{ti}^H}\text{diag}({\mathbf{g}}_{is}))\otimes\
((\beta_{s})^2\text{diag}({\mathbf{g}}_{is}^H){\mathbf{G}}_{ir}{\mathbf{G}}_{ir}^H\text{diag}({\mathbf{g}}_{is})) $
and $\hat{\mathbf{v}}_1=\text{vec}(\mathbf{v}_1\mathbf{v}_1^H)$. However, this term remains non-convex due to the quadratic forms involved. To solve this problem, the SCA method which handles \eqref{30} by its first-order Taylor expansion \cite{lipp2016variations}, is applied. Specifically, the following surrogate function is considered:
\begin{equation}
\begin{aligned}
&\hat{{\mathbf{v}}}_\mathrm{{1}}^{H}{\mathbf{P}}{\hat{{\mathbf{v}}}_\mathrm{{1}}} \approx {\left({\frac{{\partial \left({\hat{{\mathbf{v}}}_\mathrm{{1}}^{H}{\mathbf{P}}{\hat{{\mathbf{v}}}_\mathrm{{1}}}} \right)}}{{\partial {\hat{{\mathbf{v}}}_\mathrm{{1}}}}}{\bigg |_{{\hat{{\mathbf{v}}}_\mathrm{{1}}} = {\hat{\mathbf{v}}_\mathrm{{1}}^{(l)}}}}} \right)^{T}}\left({{\hat{{\mathbf{v}}}_\mathrm{{1}}} - {\hat{\mathbf{v}}_\mathrm{{1}}^{(l)}}} \right) \\ &\;\;\;\;\;\;\;\;\;\;\;\;\;\;\;\;+ {\left({\hat{\mathbf{v}}_\mathrm{{1}}^{(l)}} \right)^{H}}{\mathbf{P}}{\hat{\mathbf{v}}_\mathrm{{1}}^{(l)}},  \label{31}
\end{aligned}
\end{equation}
where $\hat{{\mathbf{v}}}_\mathrm{{1}}^{(l)}$ represents a feasible point at the $l$th iteration. Note that after each iteration, $\hat{{\mathbf{v}}}_\mathrm{{1}}^{(l)}$ will be updated according to the optimal solutions obtained.

Given that $\hat{\mathbf{v}}_{\mathrm{1}}$ and $\mathbf{P}$ are complex, let $\hat{\mathbf{v}}_{\mathrm{1}} = \Re(\hat{\mathbf{v}}_{\mathrm{1}}) + j\Im(\hat{\mathbf{v}}_{\mathrm{1}})$ and $\mathbf{P} = \Re(\mathbf{P}) + j\Im(\mathbf{P})$, where $\Im$ and $\Re$ represent the imaginary part and the real part, respectively.
Based on \cite{petersen2008matrix}, we can obtain the expression below after some matrix operations:
\begin{equation}
\begin{aligned}
&\hat{{\mathbf{v}}}_\mathrm{{1}}^{H}{\mathbf{P}}{\hat{{\mathbf{v}}}_\mathrm{{1}}}=
\Re \left({\hat{{\mathbf{v}}}_\mathrm{{1}}^{T}} \right)\Re \left({\mathbf{P}} \right)\Re \left({{\hat{{\mathbf{v}}}_\mathrm{{1}}}} \right) + \Im \left({\hat{{\mathbf{v}}}_\mathrm{{1}}^{T}} \right)\Im \left({\mathbf{P}} \right)\Re \left({{\hat{{\mathbf{v}}}_\mathrm{{1}}}} \right) \\ &\;\;\;\;\;- \Re \left({\hat{{\mathbf{v}}}_\mathrm{{1}}^{T}} \right)\Im \left({\mathbf{P}} \right)\Im \left({{\hat{{\mathbf{v}}}_\mathrm{{1}}}} \right) + \Im \left({\hat{{\mathbf{v}}}_\mathrm{{1}}^{T}} \right)\Re \left({\mathbf{P}} \right)\Im \left({{\hat{{\mathbf{v}}}_\mathrm{{1}}}} \right) \\ &\;\;\;\;\;- \Im \left({\hat{{\mathbf{v}}}_\mathrm{{1}}^{T}} \right)\Im \left({\mathbf{P}} \right)\Im \left({{\hat{{\mathbf{v}}}_\mathrm{{1}}}} \right) - \Re \left({\hat{{\mathbf{v}}}_\mathrm{{1}}^{T}} \right)\Im \left({\mathbf{P}} \right)\Re \left({{\hat{{\mathbf{v}}}_\mathrm{{1}}}} \right).
\end{aligned}
\end{equation}

Then, it can be transformed as
\begin{equation}
\begin{aligned}
\hat{{\mathbf{v}}}_\mathrm{{1}}^{H}{\mathbf{P}}{\hat{{\mathbf{v}}}_\mathrm{{1}}} &= \!\left[ \!{\Re \left({\hat{{\mathbf{v}}}_\mathrm{{1}}^{T}} \right),\Im \left({\hat{{\mathbf{v}}}_\mathrm{{1}}^{T}} \right)} \!\right]\left[ \!\begin{array}{l}\Re \left({\mathbf{P}} \right), - \Im \left({\mathbf{P}} \right)\\ \Im \left({\mathbf{P}} \right),\Re \left({\mathbf{P}} \right) \end{array} \right]\left[ \begin{array}{l}\Re \left({{\hat{{\mathbf{v}}}_\mathrm{{1}}}} \right)\\ \Im \left({{\hat{{\mathbf{v}}}_\mathrm{{1}}}} \right) \end{array} \!\right] \\ &= \tilde{{\mathbf{v}}}_\mathrm{{1}}^{T}\widetilde{{\mathbf{P}}}{\tilde{{\mathbf{v}}}_\mathrm{{1}}},\label{111}
\end{aligned}
\end{equation}
where $\tilde{{\mathbf{v}}}_\mathrm{{1}} = \left[ \begin{array}{l}\Re \left({{\hat{{\mathbf{v}}}_\mathrm{{1}}}} \right)\\ \Im \left({{\hat{{\mathbf{v}}}_\mathrm{{1}}}} \right) \end{array} \!\right]$ and $\tilde{{\mathbf{P}}}=\left[ \!\begin{array}{l}\Re \left({\mathbf{P}} \right), - \Im \left({\mathbf{P}} \right)\\ \Im \left({\mathbf{P}} \right),\Re \left({\mathbf{P}} \right) \end{array} \right]$. After differentiating \eqref{111}, we can obtain
\begin{equation}
\begin{aligned}
\frac{{{\partial \left({\tilde{{\mathbf{v}}}_\mathrm{{1}}^{T}\widetilde{{\mathbf{P}}}{\tilde{{\mathbf{v}}}_\mathrm{{1}}}} \right)}}}{{{\partial {\tilde{{\mathbf{v}}}_\mathrm{{1}}}}}}
= 2\widetilde{{\mathbf{P}}}{\tilde{{\mathbf{v}}}_\mathrm{{1}}}.\label{34}
\end{aligned}
\end{equation}

Substituting \eqref{34} in \eqref{31}, we can obtain
\begin{equation}
\begin{aligned}
\hat{{\mathbf{v}}}_\mathrm{{1}}^{H}{\mathbf{P}}{\hat{{\mathbf{v}}}_\mathrm{{1}}} \approx 2{\left({\tilde{\mathbf{v}}_\mathrm{{1}}^{(l)}} \right)^{T}}\widetilde{{\mathbf{P}}}{\tilde{{\mathbf{v}}}_\mathrm{{1}}} - {\left({\tilde{\mathbf{v}}_\mathrm{{1}}^{(l)}} \right)^{T}}\widetilde{{\mathbf{P}}}\tilde{\mathbf{v}}_\mathrm{{1}}^{(l)}.
\end{aligned}
\end{equation}

 Define $\mathbf{p}^T ={\left({\tilde{\mathbf{v}}_\mathrm{{1}}^{(l)}} \right)^{T}}\widetilde{{\mathbf{P}}}$. After several mathematical operations, we can obtain 
\begin{equation}
\begin{aligned}
{\left({\tilde{\mathbf{v}}_{\mathrm{{1}}}^{(l)}} \right)^{T}}\widetilde{{\mathbf{P}}}{\tilde{{\mathbf{v}}}_{\mathrm{{1}}}} = \Re \left({{\overline{\mathbf{p}}^{H}}{\hat{{\mathbf{v}}}_{\mathrm{{1}}}}} \right), \label{36}
\end{aligned}
\end{equation}
where ${\overline{\mathbf{p}}^{H}}={\mathbf{p}}^{T}_{1:L^2}-j{\mathbf{p}}^{T}_{L^2+1:2L^2}$, which represents the process of converting a vector ${\mathbf{p}}^{T}$ with a dimension of $2L^2\times 1$ to a vector ${\overline{\mathbf{p}}^{H}}$ with a dimension of $ L^2 \times 1$. 

Then, based on the principle of matrix calculation, \eqref{36} can be  reformulated as 
\begin{equation}
\begin{aligned}
\Re \left({{\overline{\mathbf{p}}^{H}}{\hat{{\mathbf{v}}}_{\mathrm{{1}}}}} \right) = \Re \left({{{\mathbf{v}}}_{\mathrm{{1}}}}^H{{(\sum\overline{\mathbf{p}}}){{{\mathbf{v}}}_{\mathrm{{1}}}}} \right),
\end{aligned}
\end{equation}
where $\sum\overline{\mathbf{p}}$ denotes the process of transforming a vector ${\overline{\mathbf{p}}^{H}}$ with a dimension of $ L^2 \times 1$ to a matrix with a dimension of $ L \times L$. Therefore, \eqref{A} can be reformulated as 
\begin{equation}
\begin{aligned}
 \mathbf A = 2\Re \left({{{\mathbf{v}}}_{\mathrm{{1}}}}^H{{(\sum\overline{\mathbf{p}}}){{{\mathbf{v}}}_{\mathrm{{1}}}}} \right) -{\left({\tilde{\mathbf{v}}_\mathrm{{1}}^{(l)}} \right)^{T}}\widetilde{{\mathbf{P}}}\tilde{\mathbf{v}}_\mathrm{{1}}^{(l)}.
\end{aligned}
\end{equation}

Similar to the method used in \eqref{23b}, (43) can be rewritten as
\begin{equation}
\begin{aligned}
\mathbf A = 2\text{Tr}(\mathbf F \mathbf{V}_1) -{\left({\tilde{\mathbf{v}}_\mathrm{{1}}^{(l)}} \right)^{T}}\widetilde{{\mathbf{P}}}\tilde{\mathbf{v}}_\mathrm{{1}}^{(l)},\label{fina}
\end{aligned}
\end{equation}
where $ \mathbf F = 
\begin{bmatrix}
\sum\overline{\mathbf{p}} & 0 \\
0 & 0
\end{bmatrix}$.

For \eqref{C}, let $\mathbf{I} =(\text{diag}({\mathbf{g}}_{i,q}^H){{\mathbf{G}}_{ti}}\mathbf{w}\mathbf{w}^H{{\mathbf{G}}_{ti}^H}\text{diag}({\mathbf{g}}_{i,q}))\otimes\
((\beta_{s})^2\text{diag}({\mathbf{g}}_{i,q}^H){\mathbf{G}}_{ir}{\mathbf{G}}_{ir}^H\text{diag}({\mathbf{g}}_{i,q})) $, it can be reformulated as $\tilde{{\mathbf{I}}}=\left[ \!\begin{array}{l}\Re \left({\mathbf{I}} \right), - \Im \left({\mathbf{I}} \right)\\ \Im \left({\mathbf{I}} \right),\Re \left({\mathbf{I}} \right) \end{array} \right]$, define $\mathbf{i}^T ={\left({\tilde{\mathbf{v}}_\mathrm{{1}}^{(l)}} \right)^{T}}\widetilde{{\mathbf{I}}}$ and ${\overline{\mathbf{i}}^{H}}={\mathbf{i}}^{T}_{1:L^2}-j{\mathbf{i}}^{T}_{L^2+1:2L^2}$.
Thus, it can be transformed as 
\begin{equation}
\begin{aligned}
\mathbf C= 2\text{Tr}(\mathbf L \mathbf{V}_1) -{\left({\tilde{\mathbf{v}}_\mathrm{{1}}^{(l)}} \right)^{T}}\widetilde{{\mathbf{I}}}\tilde{\mathbf{v}}_\mathrm{{1}}^{(l)},\label{fina2}
\end{aligned}
\end{equation}
where $\mathbf L  = 
\begin{bmatrix}
\sum\overline{\mathbf{i}} & 0 \\
0 & 0
\end{bmatrix} $.

Finally, the Dinkelbach method is applied for \eqref{bianhua1} to rewrite the objective function as 
\begin{equation}
\begin{aligned}
 \mathbf A -\Gamma^{(l)}({\mathbf B +\mathbf C + {\sigma _s^2}} ),\label{SS}
 \end{aligned}
\end{equation}
which is convex with respect to $\mathbf{V}_1$, where $\Gamma^{(l)}$ denotes the value of $y$ at the $l$th iteration
\begin{equation}
\begin{aligned}
\Gamma^{(l)}=\left( \frac{\mathbf A}  {\mathbf B +\mathbf C + {\sigma _s^2}}\right)^{(l-1)}.
\end{aligned}
\end{equation}

Therefore, by employing the SDR technique to omit the rank-one constraint, the sub-problem \eqref{P2_OF} transforms into 
\begin{small}\begin{align}
&\mathop {\max }\limits_{\mathbf{V}_1,\mathbf{v}_2 }\ \ \eqref{SS} \label{P2_OF2}\\
{\text{s.t.}}
&\ \ \mathbf{V}_1(m,m) + \left(\mathbf{v}_2(m)\mathbf{v}_2^H(m)\right)^T\leq 1, \forall m \in M, \tag{\ref{P2_OF2}{a}} \label{OF3e}\\
& \ \ \mathbf{V}_1 \succeq 0,\tag{\ref{P2_OF2}{b}} \\
& \ \ \eqref{SINR_K2} ,\tag{\ref{P2_OF2}{c}} \label{OF3f}
\end{align}\end{small}

\noindent which is convex and the corresponding optimal solutions can be obtained by CVX. We consider that the optimal $\overline{\mathbf v}_1^*$ can be recovered by eigenvalue decomposition if the obtained solution $\mathbf{V}_1^*$ satisfies the rank-one constraint. Otherwise, we apply the Gaussian randomization to recover $\mathbf{v}_1^*$ \cite{8811733}. Then, we can extract the phase shift vector $\mathbf{v}_1^*$ from $\overline{\mathbf v}_1^*$. Finally, $\mathbf{\Theta}_b^* = \text{diag}(\mathbf{v}_1^*)$ and $\mathbf{\Phi}_b^* = \text{diag}(\mathbf{v}_2^*)$ are obtained, respectively.

\subsection{Overall Algorithm and complexity analysis}
According to the above statement, we develop an overall iterative algorithm, denoted as Algorithm 1, for solving  \eqref{P1_OF}. The proposed algorithm is guaranteed to converge to a point because the constraints ensure that the objective function in \eqref{P1_OF} is finite. Moreover, since each sub-problem is solved either locally or optimally, the objective value of problem \eqref{P1_OF} typically improves or remains unchanged with each iteration.
\begin{algorithm}[t]
\caption{Proposed Algorithm for Solving Problem \eqref{P1_OF}. }

\begin{algorithmic}[1]
    \STATE Initialize the iteration index $n = 0$, $\{\mathbf{\Phi}_b^{(0)}, \mathbf{\Theta}_b^{(0)}\}$ and the tolerance $\delta = 10^{-4}$.
   \REPEAT 
   \STATE Initialize $i = 0,$ feasible $ y^{(0)}$ and the tolerance $\epsilon = 10^{-2}$.
    \REPEAT 
        \STATE Obtain $\{ \mathbf{w}_1^{(i)}, ..., \mathbf{w}_K^{(i)}, \mathbf{w}_s^{(i)}\}$ by solving the problem \eqref{P1_OF3} with  $y^{(i)}$.
        \STATE Update $y^{(i+1)}$ based on (21).
        \STATE Set $i = i + 1$.
    \UNTIL $\frac{y^{(i+1)}-y^{(i)}}{y^{(i)}} < \epsilon$.
   
    \STATE Initialize  $j = 0$, and  feasible $\Gamma^{(0)}$.
        \REPEAT 
        \STATE Obtain  $\{\mathbf{\Phi}_b^{(j)}, \mathbf{\Theta}_b^{(j)}\}$  by solving the problem \eqref{P2_OF2} with $\Gamma^{(j)}$.
        \STATE Update $\Gamma^{(j+1)}$ based on (47).
        \ Update $\hat{{\mathbf{v}}}_\mathrm{{1}}^{(j+1)}$, set $j = j + 1$.
    \UNTIL $\frac{\Gamma^{(j+1)}-\Gamma^{(j)}}{\Gamma^{(j)}}<\epsilon$.

\STATE Set $n = n + 1$.
\UNTIL the fractional increment of the objective function is less than $\delta$.
\end{algorithmic}
\end{algorithm}

The computational complexities of Algorithm 1 primarily arise from solving problems \eqref{P1_OF3} and \eqref{P2_OF2}.
According to \cite{qqq1}, the computational complexity of Algorithm 1 is determined by the amount of semi-definite cone constraints, the dimension of each semi-definite constraint, and the precision of addressing the SDP. Therefore, the complexity of solving \eqref{P1_OF3} is $\mathcal{O}\Big(((2K+3)(M + 1)^{\frac{7}{2}} + (2K+3)^2(M + 1)^{\frac{5}{2}} + (2K+3)^3(M + 1)^{\frac{1}{2}})l_1 \log (\frac{1}{\epsilon})\Big)$, where $l_1$ is the iteration number for optimizing the transmit beamforming. Similarly, the computational complexity of solving  \eqref{P2_OF2} is $\mathcal{O}\Big(((K+3)(L + 2)^{\frac{7}{2}} + (K+3)^2(L + 2)^{\frac{5}{2}} + (K+3)^3(L + 2)^{\frac{1}{2}})l_2 \log (\frac{1}{\epsilon})\Big)$, where $l_2$ is the iteration number for optimizing the STAR-RIS coefficients. Thus, the overall computational complexity order of the proposed algorithm is $\mathcal{O}\Big(\big((2K+3)(M + 1)^{\frac{7}{2}})l_1 \log (\frac{1}{\epsilon})+        (K+3)(L + 2)^{\frac{7}{2}})l_2 \log (\frac{1}{\epsilon})\big)l_3\Big)$,  
where the iteration number of the proposed algorithm is represented by $l_3$.

\section{Joint Beamforming and Reflecting Design for Sum Rate Maximization}
We handle (14) in this section. Similar to (13), (14) is 
an optimization problem that is both non-convex and NP-hard due to the coupling of multiple variables, making direct solutions challenging. Therefore, we divide it
into two sub-problems: the optimization of the ISAC beamforming vector, and the optimization of the STAR-RIS 
coefficients.
\subsection{ Transmit Beamforming Optimization}
In this subsection, the sub-problem of transmit beamforming defined in problem \eqref{P2} is handled, we consider that STAR-RIS coefficients $\mathbf{\Phi}_b$ and $\mathbf{\Theta}_b$ are given.
Similar to the previous section,  by applying SDR and omitting constant terms, the sub-problem of transmit beamforming can be rewritten as
\vspace{0.02em}
\begin{small}\begin{align}
\mathop {\max }\limits_{\{\mathbf{W}_k\}_{k=1}^{K},\mathbf{W}_s}&\ \ \sum\limits_{k = 1}^K {\log_2 (1+\frac{\xi_k}{\psi_k})}  \label{sumb}\\
{\text{s.t.}}\ \ & \ \   {\text{Tr}}\left( {{\sum _{k=1}^{K}\mathbf{W}_k}}+ {\mathbf{W}_s} \right) \leq P_t, \tag{\ref{sumb}{a}}   \\
& \ \  \frac{\text{Tr}\left(\mathbf{P}_s\mathbf{W}\mathbf{P}_s^H\right)}{{\text{Tr}{\left({{\mathbf{G}}_{SI}\mathbf{W}{{\mathbf{G}}_{SI}^H}}\right)}} +\text{Tr}\left({{\mathbf{P}}_q\mathbf{W}{{\mathbf{P}}_q^H}}\right) + {\sigma _s^2}} \geq \gamma_{min}. \tag{\ref{sumb}{b}} \\
& \ \ \mathbf{W}_k \succeq 0, \forall k,    \hspace{1cm}   \mathbf{W}_s \succeq 0, \tag{\ref{sumb}{c}}
\end{align}\end{small}

\noindent where $\xi_k = \text{Tr}\left({\mathbf{G}}_{t,k}\mathbf{W}_{k}{{\mathbf{G}}_{t,k}^H}\right)$ and $\psi_k={\sum\limits_{j \neq k}^K\text{Tr}{\left({{\mathbf{G}}_{t,k}\mathbf{W}_{j}{{\mathbf{G}}_{t,k}^H}}\right)}} +\text{Tr}\left({{\mathbf{G}}_{t,k}\mathbf{w}_{s}{{\mathbf{G}}_{t,k}^H}}\right) + {\sigma _k^2}$.

To tackle the non-concave objective function, we apply the inequality in (72) of \cite{budengshi} to rewrite the expression as
\vspace{0.02em}
\begin{equation}\begin{aligned} & \ln \left ({{1 + \frac {\xi_k}{\psi_k} }}\right ) \\ & \geq\ln \left ({{1 + \frac {\xi_k^{(n)}}{\psi_k^{(n)}} }}\right ) + \frac { \frac {\xi_k^{(n)}}{\psi_k^{(n)}}}{1+ \frac {\xi_k^{(n)}}{\psi_k^{(n)}}} \left ({{2 - \frac {\xi_k^{(n)}}{\xi_k} - \frac {\psi_k}{\psi_k^{(n)}} }}\right )\label{budengshi} ,\end{aligned}\end{equation}
where $\xi_k^{(n)}$ and $\psi_k^{(n)}$ are the values of $\xi_k$ and $\psi_k$ obtained in last iteration, respectively.
Then, the objective function can be reformulated as 
\begin{equation}\begin{aligned}
\log_2 \left ({{1 + \frac {\xi_k^{(n)}}{\psi_k^{(n)}} }}\right ) + \frac { \frac {\xi_k^{(n)}}{\psi_k^{(n)}}}{\log_2(1+ \frac {\xi_k^{(n)}}{\psi_k^{(n)}})} \left ({{2 - \frac {\xi_k^{(n)}}{\xi_k} - \frac {\psi_k}{\psi_k^{(n)}} }}\right ).
\end{aligned}\end{equation}
\begin{small}\begin{align}
\mathop {\max }\limits_{\{\mathbf{W}_k\}_{k=1}^{K},\mathbf{W}_s}\ \ & \text{(51)} \label{sumb1}\\
{\text{s.t.}}\ \ & \ \  \text{(49a), (49b), (49c)}, \tag{\ref{sumb1}{a}}  
\end{align}\end{small}

\noindent which is convex and the corresponding optimal solutions  $\{\mathbf{\{{W}}^*_k\}_{k=1}^{K},\{\mathbf{W}^*_s\}\}$ can
be obtained by CVX.  Based on the rank reduction theorem in Theorem 3.2 of \cite{rank1}, \eqref{sumb1} has an optimal solution $\{\mathbf{\{{W}}^*_k\}_{k=1}^{K},\{\mathbf{W}^*_s\}\}$  satisfying the rank-one constraint. The eigenvalue decomposition is applied to obtain $\{{\mathbf{\{{w}}^*_k}\}_{k=1}^{K},\{\mathbf{w}^*_s\}\}$.

\subsection{Optimization of the Coefficients for the STAR-RIS }
In this subsection, we optimize the STAR-RIS coefficients $\mathbf{\Phi}_b$ and $\mathbf{\Theta}_b$ when the beamforming matrices ${\mathbf{\{{w}}_k}\}_{k=1}^{K}$ and $\{\mathbf{w}_s\}$ are fixed. The coefficients optimization sub-problem can be expressed as
\begin{small}\begin{align}
\mathop {\max }\limits_{\mathbf{\Phi}_b,\mathbf{\Theta}_b}&\ \ \sum\limits_{k = 1}^K {\log_2 (1+\gamma_{k})}  \label{sumsum}\\
{\text{s.t.}}\ \ & \ \   \gamma_s  \geq \gamma_{min}, \tag{\ref{sumsum}{a}}   \\
& \ \  \eqref{1a}-\eqref{1c}.    \tag{\ref{sumsum}{b}} \end{align}\end{small} 

Similar with the previous section,  new variables $\mathbf{v}_1=[{{v_{b,1}}{e^{j{\nu _{b,1}}}},{v_{b,2}}{e^{j{\nu _{b,2}}}}, \ldots ,{v_{b,L}}{e^{j{\nu _{b,L}}}}}]^H$ and  $\mathbf{v}_2 =[{{u_{b,1}}{e^{j{\mu _{b,1}}}},{u_{b,2}}{e^{j{\mu _{b,2}}}}, \ldots ,{u_{b,L}}{e^{j{\mu _{b,L}}}}}] ^T$ are introduced, define $
\overline{\mathbf v}_1 =\begin{bmatrix}
\mathbf v_1 \\
1
\end{bmatrix}$, $\mathbf{V}_1 =\overline{\mathbf v}_1 \overline{\mathbf v}_1^H$, where $\mathbf{V}_1\succeq 0$ and $\text{rank}(\mathbf{V}_1)=1$. 

The approximations and matrix operations that apply in the previous section are still useful here to tackle the non-convex terms. Applying \eqref{SI}, \eqref{fina} and \eqref{fina2} to \text{(52a)}, yields 
\begin{equation}
\begin{aligned}
\mathbf A \geq \gamma_{min}(\mathbf B +\mathbf C + \sigma _s^2).
\end{aligned}
\end{equation}

Next, the non-convex objective function is handled similar with \eqref{SINR_K2}, the term can be reformulated as $ \sum\limits_{k = 1}^K {\log_2 (1+\frac{\chi_k}{\zeta_k})}$, where $\chi_k = \text{Tr}(\mathbf{U}_{k,k}{\mathbf V}_1)$ and $\zeta_k = \sum\limits_{j \neq k}^K \text{Tr}(\mathbf{U}_{k,j}{\mathbf V}_1) + \text{Tr}(\mathbf{U}_{k,s}{\mathbf V}_1) + {\sigma _k^2} $. According to \eqref{budengshi}, the objective function can be reformulated as
\begin{small}\begin{align}
\sum\limits_{k = 1}^K \left (\log_2 \left ({{1 + \frac {\chi_k^{(n)}}{\zeta_k^{(n)}} }}\right ) + \frac { \frac {\chi_k^{(n)}}{\zeta_k^{(n)}}}{\log_2(1+ \frac {\chi_k^{(n)}}{\zeta_k^{(n)}})} \left ({{2 - \frac {\chi_k^{(n)}}{\chi_k} - \frac {\zeta_k}{\zeta_k^{(n)}} }}\right )\right ) , 
\end{align}\end{small} 
where $\chi_k^{(n)}$ and $\zeta_k^{(n)}$ are the value of $\chi_k$ and $\zeta_k$ obtained in last iteration, respectively.
As a result, by employing the SDR technique to omit the rank-one constraint, the original problem \eqref{sumsum} can be restructured as follows:
\begin{small}\begin{align}
\mathop {\max }\limits_{\mathbf{\Phi}_b,\mathbf{\Theta}_b}&\ \ (55) \label{last}\\
{\text{s.t.}}\ \ 
& \ \   \mathbf{V}_1(m,m) + \left(\mathbf{v}_2(m)\mathbf{v}_2^H(m)\right)^T\leq 1, \forall m \in M,     \tag{\ref{last}{a}}\\
& \ \ \mathbf{V}_1 \succeq 0,\tag{\ref{last}{b}} \\
& \ \ (54),\tag{\ref{last}{c}} \end{align}\end{small} 

\noindent which is convex and the corresponding optimal solutions can be obtained by CVX. We consider that the optimal $\overline{\mathbf v}_1^*$ can be recovered by eigenvalue decomposition if the obtained solution $\mathbf{V}_1^*$ satisfies the rank-one constraint. Otherwise, we apply the Gaussian randomization to recover $\mathbf{v}_1^*$ \cite{8811733}. Then, we can extract the phase shift vector $\mathbf{v}_1^*$ from $\overline{\mathbf v}_1^*$. Finally, $\mathbf{\Theta}_b^* = \text{diag}(\mathbf{v}_1^*)$ and $\mathbf{\Phi}_b^* = \text{diag}(\mathbf{v}_2^*)$ are obtained, respectively.

\subsection{Overall Algorithm and complexity analysis}
According to the above statement, we develop an overall iterative algorithm Algorithm 2 for solving  \eqref{P2}. The proposed algorithm is guaranteed to converge to a point because the constraints ensure that the objective function in \eqref{P2} is finite. Moreover, since each sub-problem is solved either with a local or optimal solution, the objective value of problem \eqref{P2} typically improves or remains unchanged with each iteration.
\begin{algorithm}[t]
\caption{Proposed Algorithm for Solving Problem \eqref{P2}. }
\begin{algorithmic}[1]
    \STATE Initialize the iteration index $n = 0$, $\{\mathbf{\Phi}_b^{(0)}, \mathbf{\Theta}_b^{(0)}\}$ and the tolerance $\delta = 10^{-4}$.
   \REPEAT 
   \STATE Initialize $i = 0,$ feasible $ \xi_k^{(0)}$, $\psi_k^{(0)}$ and the tolerance $\epsilon = 10^{-2}$.
    \REPEAT 
        \STATE Obtain $\{ \mathbf{w}_1^{(i)}, ..., \mathbf{w}_K^{(i)}, \mathbf{w}_s^{(i)}\}$ by solving the problem \eqref{sumb1} with $ \xi_k^{(i)}$ and $\psi_k^{(i)}$.
        \STATE Update $ \xi_k^{(i+1)}$ and $\psi_k^{(i+1)}$.
        \STATE Set $i = i + 1$.
    \UNTIL $\frac{\xi_k^{(i+1)}-\xi_k^{(i)}}{\xi_k^{(i)}}<\epsilon $ and $\frac{\psi_k^{(i+1)}-\psi_k^{(i)}}{\psi_k^{(i)}}<\epsilon $
   
    \STATE Initialize  $j = 0$, and  feasible $\chi_k^{(0)}$ and $\zeta_k^{(0)}$ .
        \REPEAT 
        \STATE Obtain $\{\mathbf{\Phi}_b^{(j)}, \mathbf{\Theta}_b^{(j)}\}$ by solving the problem \eqref{last} with $\chi_k^{(j)}$ and $\zeta_k^{(j)}$ .
        \STATE Update $\chi_k^{(j+1)}$ and $\zeta_k^{(j+1)}$.
        \STATE Update $\hat{{\mathbf{v}}}_\mathrm{{1}}^{(j+1)}$, set $j = j + 1$.
    \UNTIL $\frac{\chi_k^{(j+1)}-\chi_k^{(j)}}{\chi_k^{(j)}}<\epsilon $ and $\frac{\zeta_k^{(j+1)}-\zeta_k^{(j)}}{\zeta_k^{(j)}}<\epsilon $

\STATE Set $n = n + 1$.
\UNTIL the fractional increment of the objective function is less than $\delta$.
\end{algorithmic}
\end{algorithm}

The computational complexity of Algorithm 2 is mainly due to solving \eqref{sumb1} and \eqref{last}. Similar to Algorithm 1, as indicated in \cite{qqq1}, the computational complexity of Algorithm 2 is determined by the number of semi-definite cone constraints, the dimension of each semi-definite constraint, and the precision required for addressing the SDP. Consequently, the computational complexity of solving \eqref{sumb1} is given by $\mathcal{O}\Big(((K+4)(M + 1)^{\frac{7}{2}} + (K+4)^2(M + 1)^{\frac{5}{2}} + (K+4)^3(M + 1)^{\frac{1}{2}})l_4 \log (\frac{1}{\epsilon})\Big)$, where $l_4$ represents the number of iterations for optimizing the transmit beamforming. Similarly, the complexity of solving \eqref{P2_OF2} is $\mathcal{O}\Big((4(L + 2)^{\frac{7}{2}} + 4^2(L + 2)^{\frac{5}{2}} + 4^3(L + 2)^{\frac{1}{2}})l_5 \log (\frac{1}{\epsilon})\Big)$, where $l_5$ is the number of iterations for optimizing the STAR-RIS coefficients. Thus, the overall computational complexity of the algorithm is $\mathcal{O}\Big(\big(((K+4)(M + 1)^{\frac{7}{2}})l_4 \log (\frac{1}{\epsilon}) + 4(L + 2)^{\frac{7}{2}})l_5 \log (\frac{1}{\epsilon})\big)l_6\Big)$, where $l_6$ is the number of iterations of the proposed algorithm.

\section{Simulation Results}
We evaluate the performance of the proposed STAR-RIS-enabled FD ISAC system using computer simulations in this section. In the simulation setup, the FD BS is equipped with $M = 6$ transmit antennas and $N = 2$ receive antennas. The first transmit and receive antennas are located at [0, 0, 5] m and [0, 0.1, 5] m, respectively. The STAR-RIS with 64 elements, each spaced by half a wavelength, is deployed with the center element located at [0.5, 0, 5] m.
For communication, the FD BS serves $K = 2$ downlink single-antenna users, located at a distance of $r_{i,1} = r_{i,2} = 80$ m, with azimuth angles of $\phi_{i,1} = 20^\circ$ and $\phi_{i,2} = 150^\circ$, and elevation angles of $\theta_{{i,1}} = \theta_{i,2} = 30^\circ$. For sensing, a single target is located at a distance of $r_{is} = 4$ m, with an azimuth angle of $\phi_{is} = 90^\circ$ and an elevation angle of $\theta_{is} = 30^\circ$. An interference source is located at a distance of $r_{i,q} = 10$ m, with an azimuth angle of $\phi_{i,q} = 120^\circ$ and an elevation angle of $\theta_{i,q} = 30^\circ$. The reflection factors of the single target and interference source are set to $\beta_r = \beta_q = 0.1$.
Additionally, we utilize a frequency band of 6 GHz and set $\lambda = 0.05$ m. The channel gain, Rician factor and pathloss exponent are ${g_0} = 1 \times 10^{-4}$, $\vartheta = 3$ and $\kappa=2.5$, respectively. The maximum transmit power is $P_t = 30$ dBm, and the noise powers for communication and sensing are ${\sigma_k^2} = {\sigma_s^2} = -80$ dBm. The required SINR thresholds for sensing and communication are set to $\gamma_{req} = 15$ dB and $\gamma_{min} = 10$ dB, respectively.

To verify the performance of the proposed scheme, a benchmark for comparison is adopted as
\begin{enumerate}
\item \textbf{FD-ISAC system without STAR-RIS:} In this case, we consider an FD-ISAC system without STAR-RIS. A traditional SIC approach is applied to reduce the SI, where the residual self-interference (RSI) received by the receive antennas is proportional to the transmit power of the transmit antennas, expressed as $P_{si} = \rho P_B$ \cite{SIC}. Let $\rho = -120$ dB denote high SIC performance.
\end{enumerate}

\begin{figure}[t]
    \centering
    \includegraphics[width=0.5\textwidth]{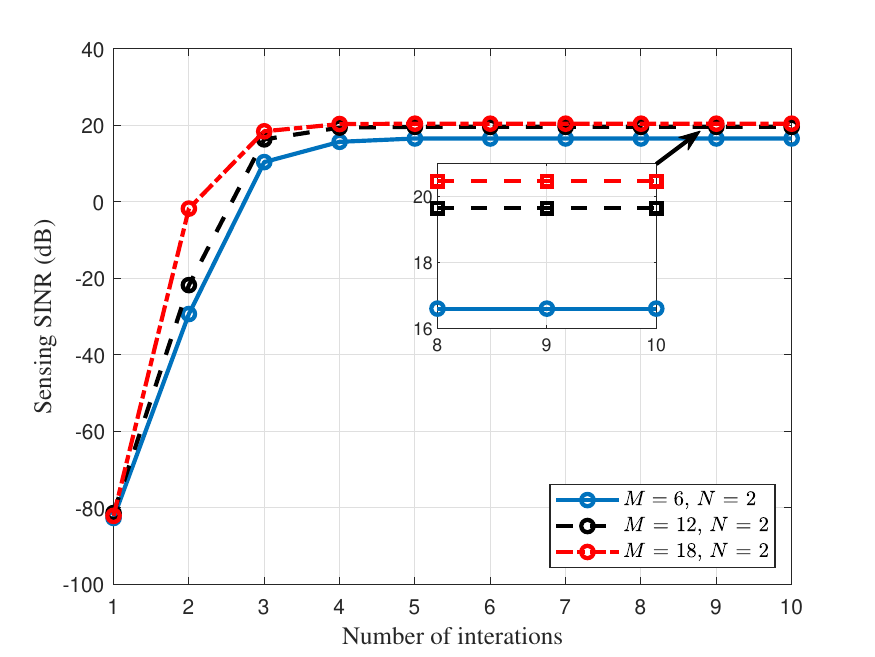}
    \caption{Validation of Convergence for Algorithm 1.}
    \label{fig:example}
\end{figure}

The convergence performance of proposed Algorithm 1 for different numbers of transmit antennas is verified in Fig. 2. From this figure, it is observed that the sensing SINR converges quickly to a stable value as the number of iterations increases, which demonstrates the feasibility of Algorithm 1. 

In what follows, the transmission beampattern gain of the STAR-RIS for sensing functionality with different minimum communication SINR thresholds $\gamma_{req}$, achieved by Algorithm 1, is shown in Fig. 3. Since the users and the detection target are located on the same horizontal line, i.e., $\theta_{i,\text{z}} = 30^{\circ}$ for $\text{z} = 1, \ldots, K, s$, we focus on the beampattern gain for different azimuths $\phi_{i,\text{z}}$ at the same elevation angle. Therefore, the beampattern map is two-dimensional. This figure shows that the transmission beamforming from the STAR-RIS to the detection target and the downlink users, i.e., $\phi_{i,\text{z}} = 0^{\circ} - 180^{\circ}$.
According to Fig. 3, the transmission beam concentrates on the detection target and the communication users, and the peak of the beam towards the target becomes lower at higher $\gamma_{req}$, while the beams towards the users show the opposite trend. It should be mentioned that since the beamforming for different communication users and the detection target is superimposed, the beam directed towards a specific user is less distinct because its wave peak is the superposition of the wave troughs of beams for other users at this angle. Meanwhile, there is a null towards the interference source, which reduces environmental clutter. Hence, the effectiveness of Algorithm 1 regarding the sensing functionality is validated.
\begin{figure}[t]
    \centering
    \includegraphics[width=0.5\textwidth]{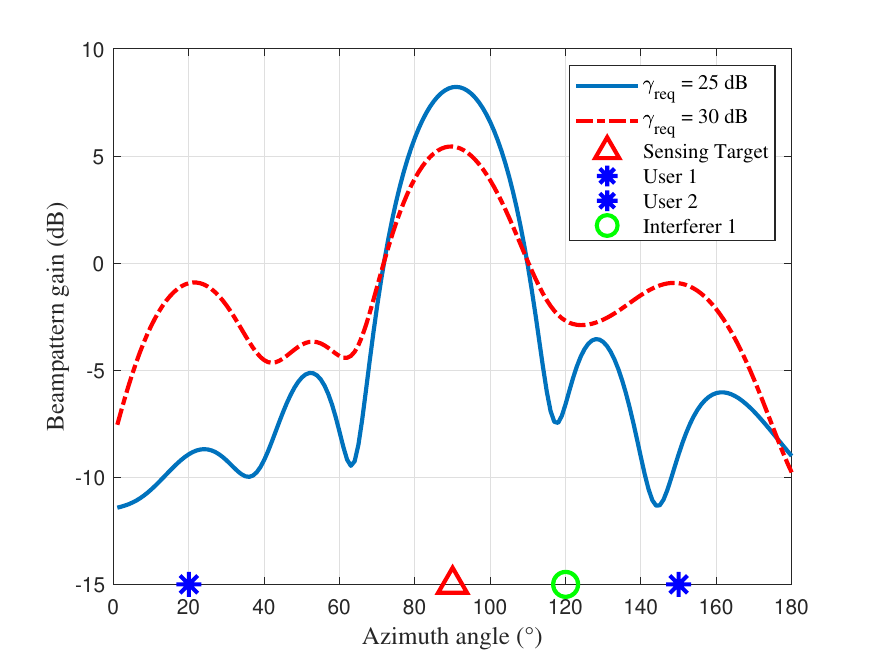}
    \caption{Transmission beampattern of the STAR-RIS regarding the sensing functionality of Algorithm 1.}
    \label{fig:example}
\end{figure}

To demonstrate the performance of reducing SI, we compared the sensing SINR with different numbers of STAR-RIS elements $L$ in Fig. 4. We observe that the sensing SINR rises steadily as the number of elements grows. This is due to a larger number of elements $L$ not only enhances the strength of the sensing and communication signals but also improves the capability of reducing SI. Furthermore, it is obvious that with the same number of transmit antennas, the system performance degrades with an increased number of receive antennas. This is because a larger number of receive antennas makes the reduction of SI more difficult. This effect is clearly demonstrated in the sensing SINR when $M = 6$ and $N = 3$, which shows a poor capability of reducing SI when the number of elements is small (from $5 \times 5$ to $7 \times 7$). Additionally, even with a larger number of elements (greater than $10 \times 10$), the results with more receive antennas do not outperform those with fewer receive antennas as expected, indicating that the functionality of reducing SI still imposes a limitation on the overall system performance.

\begin{figure}[t]
    \centering
    \includegraphics[width=0.5\textwidth]{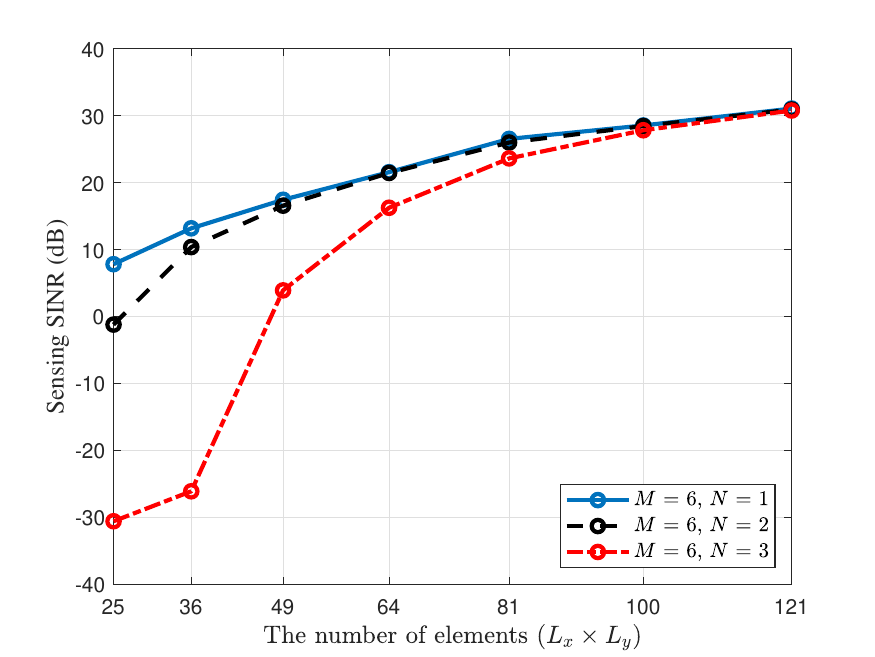}
    \caption{Sensing SINR comparison with the different number of STAR-RIS elements.}
    \label{fig:example}
\end{figure}
\begin{figure}[t]
    \centering
    \includegraphics[width=0.5\textwidth]{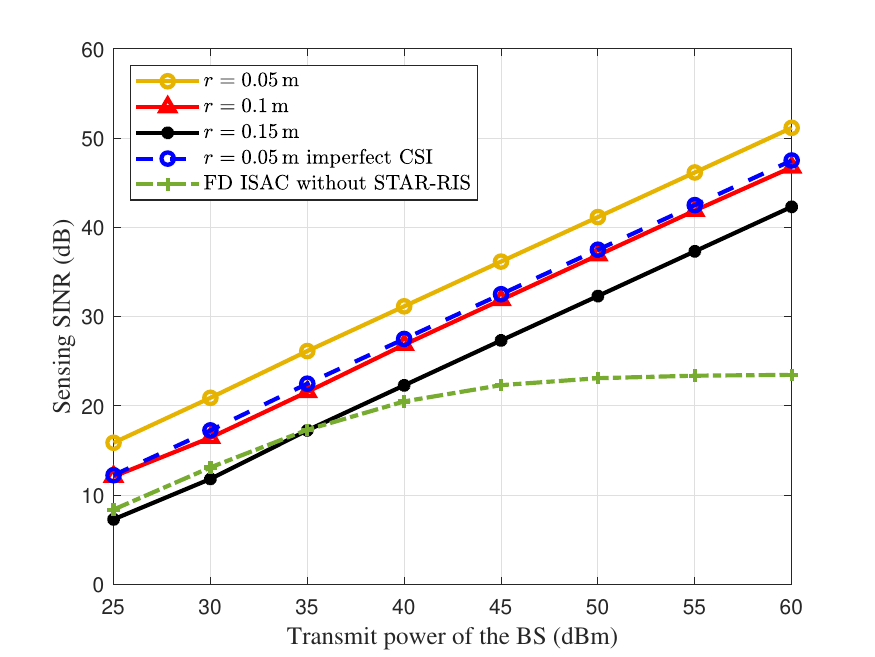}
    \caption{Sensing SINR comparison with different transmit power at FD BS.}
    \label{fig:example}
\end{figure}
Fig. 5 shows the comparison of sensing SINR with a benchmark under different transmit power levels at the FD BS.  It can be seen that as the transmit power at the FD BS increases, the system's sensing SINR gradually improves. However, when the distance between the STAR-RIS and the FD BS increases, the sensing performance deteriorates. This is because the increased distance leads to higher path loss in the STAR-RIS reflection path $\mathbf{G}_{ir}$, requiring the STAR-RIS to use higher reflection power to reduce SI, thereby decreasing overall system performance. Additionally, the increased distance also results in higher path loss between the transmit antennas and the STAR-RIS, causing significant attenuation in $\mathbf{G}{ti}$, which can degrade the performance of both sensing and communication.

The performance with imperfect CSI is also evaluated and shown in Fig. 5. The estimated channel between the STAR-RIS and the detection target is assumed to be $\hat{\mathbf{g}}_{i,\text{z}} = \sqrt{1-\epsilon} \mathbf{g}_{i,\text{z}} + \sqrt{\epsilon} \tilde{\mathbf{g}}_{i,\text{z}}$, where $\text{z}=1,\ldots,k,s$, $\epsilon \in [0, 1]$ denotes the estimation error variance, and $\tilde{\mathbf{g}}_{i,\text{z}}$ is the channel estimation error vector with $\left[ \tilde{\mathbf{g}}_{i,\text{z}} \right]_l \sim \mathcal{CN} (0, |\left[ \mathbf{g}_{i,\text{z}} \right]_l |)$, where $\text{z}=1,\ldots,k,s$. Set $\epsilon = 0.1$ according to \cite{8910627,9732214}.
It can be seen that imperfect CSI causes a degradation in sensing SINR compared to perfect CSI.

The benchmark FD-ISAC system without STAR-RIS is also shown in Fig. 5 to compare the proposed scheme with the traditional SIC scheme. It can be seen that at low power levels (below 40 dBm), the traditional SIC scheme performs well, but when the power exceeds 40 dBm, the sensing SINR of the traditional SIC scheme no longer improves. This is because, in the traditional SIC scheme, both the power of sensing echo and SI are proportional to the transmit power. When the transmit power is high, the power of SI becomes much greater than the noise, resulting in the sensing SINR being the ratio of sensing echo to the SI. Since both are proportional to the transmit power, the sensing SINR  remains unchanged. Furthermore, it is obvious that the performance of the schemes with distances of 0.05 m and 0.1 m surpass the traditional SIC scheme at any transmit power level. This indicates that the proposed scheme can effectively reduce SI and is not affected by the signal power.

\begin{figure}[t]
    \centering
    \includegraphics[width=0.5\textwidth]{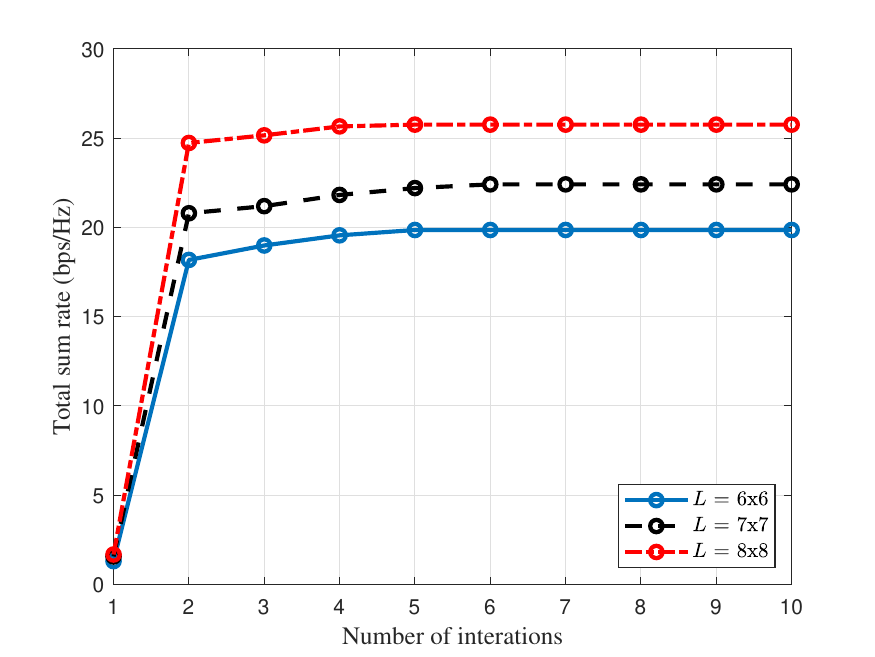}
    \caption{Convergence validation of Algorithm 2.}
    \label{fig:example}
\end{figure}
\begin{figure}[t]
    \centering
    \includegraphics[width=0.5\textwidth]{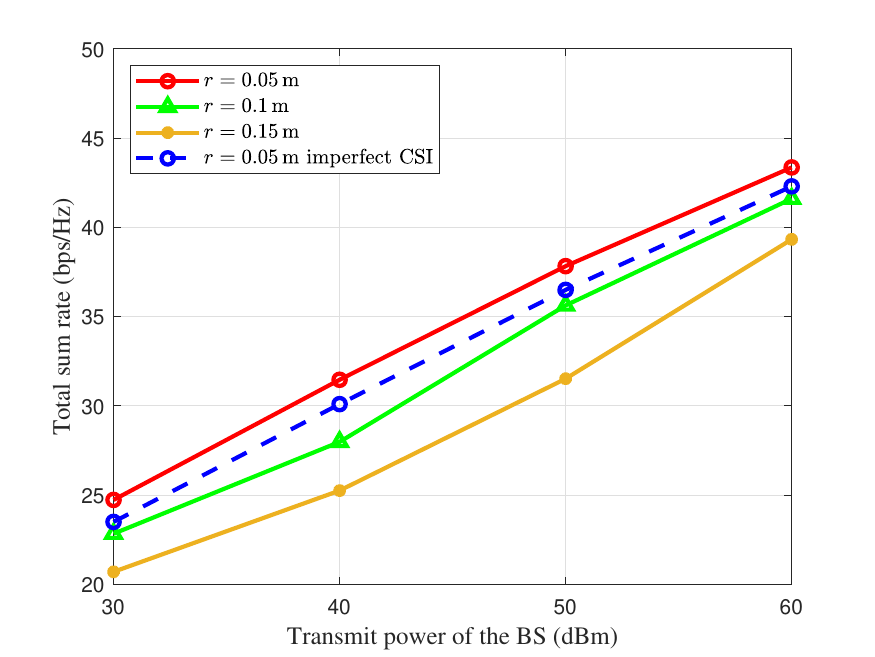}
    \caption{Total sum rate comparison with different transmit power at FD BS.}
    \label{fig:example}
\end{figure}

The convergence performance of the proposed Algorithm 2 for different numbers of STAR-RIS elements is demonstrated in Fig. 6. As shown in the figure, the total sum rate rapidly converges to a stable value as the number of iterations increases, confirming the feasibility of Algorithm 2. Furthermore, the figure indicates that a greater number of elements can not only enhance the performance of communication and sensing but also improve the capability to reduce SI.

In what follows, we compare the total sum rate with different transmit power at FD BS achieved by Algorithm 2 in Fig. 7. It is evident that as the transmit power increases, the total sum rate gradually rises. Additionally, when the distance between the STAR-RIS and the FD BS is large, the communication performance decreases. This is because the increased distance leads to higher path loss in the STAR-RIS reflection path $\mathbf{G}_{ir}$, which requires more reflection power to reduce SI, reducing overall system performance. It also increases path loss between the transmit antennas and the STAR-RIS, significantly weakening $\mathbf{G}_{ti}$ and thus impairing both sensing and communication capabilities. The performance with imperfect CSI is also compared in the figure. Same as the scheme of imperfect CSI in Fig. 5, we set $\epsilon = 0.1$. It can be seen that imperfect CSI causes a degradation in sensing SINR compared to perfect CSI.  Overall, the figure demonstrates the effectiveness of the proposed scheme in enhancing communication functionality.

\section{Conclusion}
In this paper, we proposed a new advanced STAR-RIS-enabled FD ISAC system to not only enhance simultaneous communication and target sensing but also effectively reduce SI. We investigated the joint optimization of a STAR-RIS-enabled FD-ISAC system under the criteria of sensing SINR maximization and communication sum rate maximization. To address the non-convex optimization problems with multiple coupled parameters, we proposed alternating optimization algorithms for each optimization problem. Firstly, we developed an SDR-based algorithm for transmit beamformer design. Then, an SCA scheme was adopted, and an SDR-based algorithm for refracting and reflecting coefficient design was proposed. Simulation results verified the effectiveness of deploying STAR-RIS in enhancing the performance of communication and sensing while also reducing SI. Moreover, the numerical results showed that the proposed scheme outperforms the traditional SIC scheme, especially at high transmit power.

\footnotesize
\bibliographystyle{ieeetr}
\bibliography{ref}
\end{document}